\documentclass[12pt]{article}
\usepackage[top=2.5cm, left=2.5cm, bottom=2.5cm, right=2.5cm]{geometry} 
\usepackage[english]{babel}
\usepackage[square,numbers]{natbib}
\usepackage{setspace}
\usepackage{datetime}
\usepackage{mathptmx}
\usepackage{multirow}
\usepackage{multicol}
\usepackage{graphicx}
\usepackage{subcaption}
\usepackage{siunitx}
\usepackage{microtype}
\usepackage{amsmath,amsthm,amssymb,mathtools}
\usepackage[affil-it]{authblk}
\usepackage{xurl}
\usepackage{indentfirst}
\usepackage[colorlinks = true, linkcolor = black, citecolor=black, urlcolor=blue, filecolor = black, bookmarks=false,hypertexnames=true]{hyperref}
\graphicspath{{./}}

\title{Analyzing Errors in Controlled Turret System Given Target Location Input from Artificial Intelligence Methods in Automatic Target Recognition}

\author[1]{Matthew Karlson}
\author[1]{Heng Ban \thanks{Corresponding Author. \textit{Email Address: heng.ban@pitt.edu}}}
\author[1]{Daniel G. Cole}
\author[2]{Mai Abdelhakim}
\author[3]{Jennifer Forsythe}

\affil[1]{Department of Mechanical Engineering and Materials Science \protect\\ University of Pittsburgh, Pittsburgh, PA 15260 \vspace{0.5em}}
\affil[2]{Department of Electrical and Computer Engineering \protect\\ University of Pittsburgh, Pittsburgh, PA 15260 \vspace{0.5em}}
\affil[3]{DEVCOM Analysis Center \protect\\ Aberdeen Proving Ground, MD 21005}

\date{}

\begin{document}

\maketitle
\begin{abstract}
In this paper, we assess the movement error of a targeting system given target location data from artificial intelligence (AI) methods in automatic target recognition (ATR) systems. Few studies evaluate the impacts on the accuracy in moving a targeting system to an aimpoint provided in this manner. To address this knowledge gap, we assess the performance of a controlled gun turret system given target location from an object detector developed from AI methods. In our assessment, we define a measure of object detector error and examine the correlations with several standard metrics in object detection. We then statistically analyze the object detector error data and turret movement error data acquired from controlled targeting simulations, as well as their aggregate error, to examine the impact on turret movement accuracy. Finally, we study the correlations between additional metrics and the probability of a hit. The results indicate that AI technologies are a significant source of error to targeting systems. Moreover, the results suggest that metrics such as the confidence score, intersection-over-union, average precision and average recall are predictors of accuracy against stationary targets with our system parameters.
\end{abstract}

\clearpage
\section{Introduction}
The purpose of this paper is to assess the movement errors of a targeting system given target location data from object detection models based on artificial intelligence (AI) technology that are leveraged in automatic target recognition (ATR) systems. The scope of current research does not address the impacts on the error in moving a targeting system to an aimpoint measured from the output of these models after a detection has been made. In particular, it is unclear which metrics of performance could be predictors of accuracy for targeting systems given target location input from an object detection model designed for ATR. Having reliable metrics is necessary for a human observer to weigh important outcomes depending on the veracity of ATR output, which is crucial to preserving human safety and well-being.

ATR is an important field of research with applications traditionally devoted to military purposes \citep{Bhanu1986Automatic,Gilmore1984Artificial,Sikka1989Distributed,Verly1989Machine}. More recently, however, ATR technology has been extended to applications in the civilian area \citep{Christiansen2014Automated,FernandezCaballero2014Thermal,Pierucci2007Improvements}. The goal of an ATR system is to perform real-time target detection, identification and classification by processing sensory image data; this data may come from imaging modalities such as synthetic aperture radar (SAR) \citep{Blasch2020Review,ElDarymli2016Automatic,KechagiasStamatis2021Automatic} or electro-optical (EO) and infrared (IR) \citep{Fan2019Challenges,Ratches2011} sensors. The image data is processed through  a suite of algorithms performing an array of tasks including filtering, detection, segmentation, feature selection, classification, prioritization and aimpoint selection \citep{Bhanu1986Automatic}. Research in AI methods to handle these ATR functions dates back to the 1980's \citep{Gilmore1984Artificial,Sikka1989Distributed,Verly1989Machine}.

Currently, state of the art ATR approaches are developed from deep learning (DL) architectures in the field of AI, the most popular of which is the convolutional neural network (CNN) \citep{Chen2016Target,Kazemi2018Automatic,MalmgrenHansen2015Convolutional,Pathak2018Application,Profeta2016Convolutional,Shetty2021object,Wang2015Application}. The advantage of using CNNs is that they streamline the detection process by allowing for both localization and classification of objects to be combined into one pipeline \citep{Padilla2021Comparative}. CNNs achieve this by performing the convolution operation between network parameters and subregions of the image in each convolutional layer rather than standard matrix multiplication as in a conventional feedforward neural network (FNN). This operation reduces the number of parameters that need to be determined during network training.

CNNs used for object detection in ATR are trained with supervised learning techniques similar to archetypical FNN models used for regression or classification tasks; examples of such models in the literature include OverFeat \citep{Sermanet2014OverFeat}, SPP-net \citep{He2015Spatial}, R-CNN \citep{Girshick2014Rich}, Fast R-CNN \citep{Girshick2015Fast}, Faster R-CNN \citep{Ren2017Faster}, SSD \citep{Liu2016SSD} and YOLO \citep{Redmon2016You}. The procedure in the supervised learning approach is to determine the parameters that minimize a loss function between training data and model predictions in a gradient descent algorithm that applies backpropagation and iteratively updates the parameters until convergence. Training data consists of a set of images annotated with predefined class labels of objects and bounding box data, referred to as the ground truth data. The bounding box is an annotated rectangular region in the image surrounding the object.

After training and testing, the model can be deployed for real-time object detection. When performing detection, the model takes in an image and filters it through the convolutional layers of the CNN to extract feature maps, which are regions of pixels in the image that are most representative of predefined class objects. These feature maps are concatenated into one feature vector that is passed through the fully connected layers of an FNN to perform classification and regression tasks. The final outputs of the model are predictions of bounding boxes surrounding detected objects, class labels and associated confidence scores of each detection. The bounding box data is in the same format as the ground truth data used to train the model. The confidence score is a number ranging from \num{0} to \num{1} that intends to provide a level of confidence that the detected object is a relevant target and lies within the predicted bounding box \citep{Padilla2020Survey}; the closer the number is to \num{1} the higher the confidence of the prediction. 

Performance metrics for object detectors in ATR assess both detection and classification ability. The assessment is done on a testing set of images with annotated ground truth objects not used for training the object detection algorithm.  Historically, the three main metrics have been the probability of target detection, the probability of correct classification and the false alarm rate \citep{Bhanu1986Automatic}. The probability of target detection is evaluated against the false alarm rate on a receiver operator characteristic (ROC) curve, while the probability of correct classification is determined from the confusion matrix displaying the number of correct and incorrect classifications of each target class (e.g., Figure 3 and Figure 4 in the survey by \citet{Ratches2011}). As CNNs are becoming a dominant architecture in ATR for object detection, state of the art metrics have been expanded to include the confidence score and a measure of overlap between the predicted and ground truth bounding boxes \citep{Padilla2020Survey,Padilla2021Comparative}. This overlap is quantified by the intersection-over-union (IoU), which is defined as the area of the intersection between the predicted and ground truth bounding boxes divided by the area of their union; it is based on the Jaccard index measuring the similarity between two sets of data \citep{Jaccard1901Etude}. As the IoU lies between \num{0} and \num{1}, the closer the value is to \num{1}, the better the detection performance of the ATR system. However, the IoU only indicates detection ability, it says nothing about classification ability. This has led to the development of more comprehensive metrics in object detection that measure both localization and classification performance. 

Average precision (AP) is a metric quantifying both localization as well as classification \citep{Girshick2015Fast,Girshick2014Rich,He2015Spatial,Liu2016SSD,Padilla2020Survey,Padilla2021Comparative,Redmon2016You,Ren2017Faster,Sermanet2014OverFeat}, which is defined based on the concepts of precision and recall. Precision measures the correctness of the detections made being the percentage of correct detections out of all detections at a given confidence level. Recall measures the ability of the object detector to correctly detect all ground truth objects quantifying the percentage of correct detections out of all ground truth objects at a given confidence level. The IoU is used to set a threshold indicating whether a detection is correct or incorrect – a detection with IoU above the threshold is correct, while those below the threshold are incorrect. Average recall (AR) is another metric that assesses the localization ability of the detector \citep{Hosang2016What,Padilla2020Survey}; being based on recall, it is tailored to applications that require accurate detection proposals of ground truth objects. Because both AP and AR are dependent on the IoU, there are several variants of each metric used to evaluate object detectors in ATR. For more details regarding the definitions of each metric and how they are evaluated, please see Section \ref{metricsForObjectDetectors}.

AI techniques have led to novel object detection algorithms in ATR with significant improvements in detection and classification performance. However, the research in this area has traditionally been limited to assessments of the various tasks executed by the object detector as it processes image data to detect and identify targets. In evaluation studies of state of the art object detection models \citep{Girshick2015Fast,Girshick2014Rich,He2015Spatial,Liu2016SSD,Redmon2016You,Ren2017Faster,Sermanet2014OverFeat}, the performance metrics have only been applied to evaluate the effects of these tasks on the overall output of the detector. These studies do not address the impacts on the accuracy in moving a targeting system to an aimpoint determined after target detection. On the other hand, the limited amount of research in the literature on the impacts of object detection technology in ATR addresses only the human factor effects. Specifically, these studies assess human detection and identification performance in simulated targeting environments in which ATR technology assists participants in target acquisition tasks \citep{Gardony2022Aided,Hollands2018Effects,Reiner2017Target,Riegler1998Human}. In contrast, we are interested in assessing the performance of targeting systems assisted by feedback controllers in moving to an aimpoint measured from target location data provided by an object detector.

In this paper, we assess the impacts of AI technology on the targeting performance of a controlled gun turret model given target location as input in numerical simulations. An object detection model based on a Faster R-CNN architecture provides target location of stationary targets to the controlled gun turret in the simulations. In our assessment, we first define a measure of object detection error and examine the correlations between this error and metrics such as the confidence score, IoU and detected bounding box area. We then analyze the distribution of the object detection error data, the turret movement error data acquired from the targeting simulations and their aggregate error data. Continuing our analysis, we evaluate several variants of AP and AR to determine which metrics could serve as predictors of high targeting performance given AI input of target location. In this effort, we examine the correlations of these metrics with the probability of a hit evaluated in additional targeting simulations of the controlled gun turret given target location as input. In these simulations, target location is provided to the gun turret from five datasets sampled from the output dataset of the assessed object detection model. The probability of a hit is evaluated  using the ground truth bounding box data at the end of each simulation. For each of the five datasets, we average the probability of a hit over the number of detections in the corresponding dataset. This procedure is repeated for six uniformly spaced target ranges from \SI{500}{\meter} to \SI{3000}{\meter}.

\section{Performance Metrics}
\label{methods}
\label{metricsForObjectDetectors}

\subsection{Measure of AI Error}
\label{measureOfAiError}
The metric for quantifying the error introduced to our model targeting system by the object detector is the centroid error between the ground truth bounding box and a detected bounding box in an image; we will refer to this error as the AI error in continuation. But before defining the AI error precisely, we need to introduce some notation concerning the location of pixels in an image. A vector of pixel coordinates relative to some origin $O$ in an image is denoted as $\mathbf{x}=(x,y)$, where $x$ is the horizontal coordinate and $y$ is the vertical coordinate. In this work, $O$ is either the top-left corner, bottom-left corner, or center of the image and will be specified if not clear from the context. The Euclidean length of a vector $\mathbf{x}$ is denoted $\Vert\mathbf{x}\Vert=\sqrt{x^2+y^2}$. We denote $\mathbf{x}_g=(x_g,y_g)$ as the centroid of the ground truth bounding box and $\mathbf{x}_b=(x_b,y_b)$ as the centroid of the predicted bounding box. We will refer to $\mathbf{x}_g$ as the true aimpoint and $\mathbf{x}_b$ as the ATR measured aimpoint or set point. We then define the AI error as  $\Vert\mathbf{R}_{bg}\Vert$, where $\mathbf{R}_{bg}=\mathbf{x}_b-\mathbf{x}_g$. 

\subsection{Intersection-over-Union}
\begin{figure}
        \centering
        \includegraphics[width=0.5\linewidth]{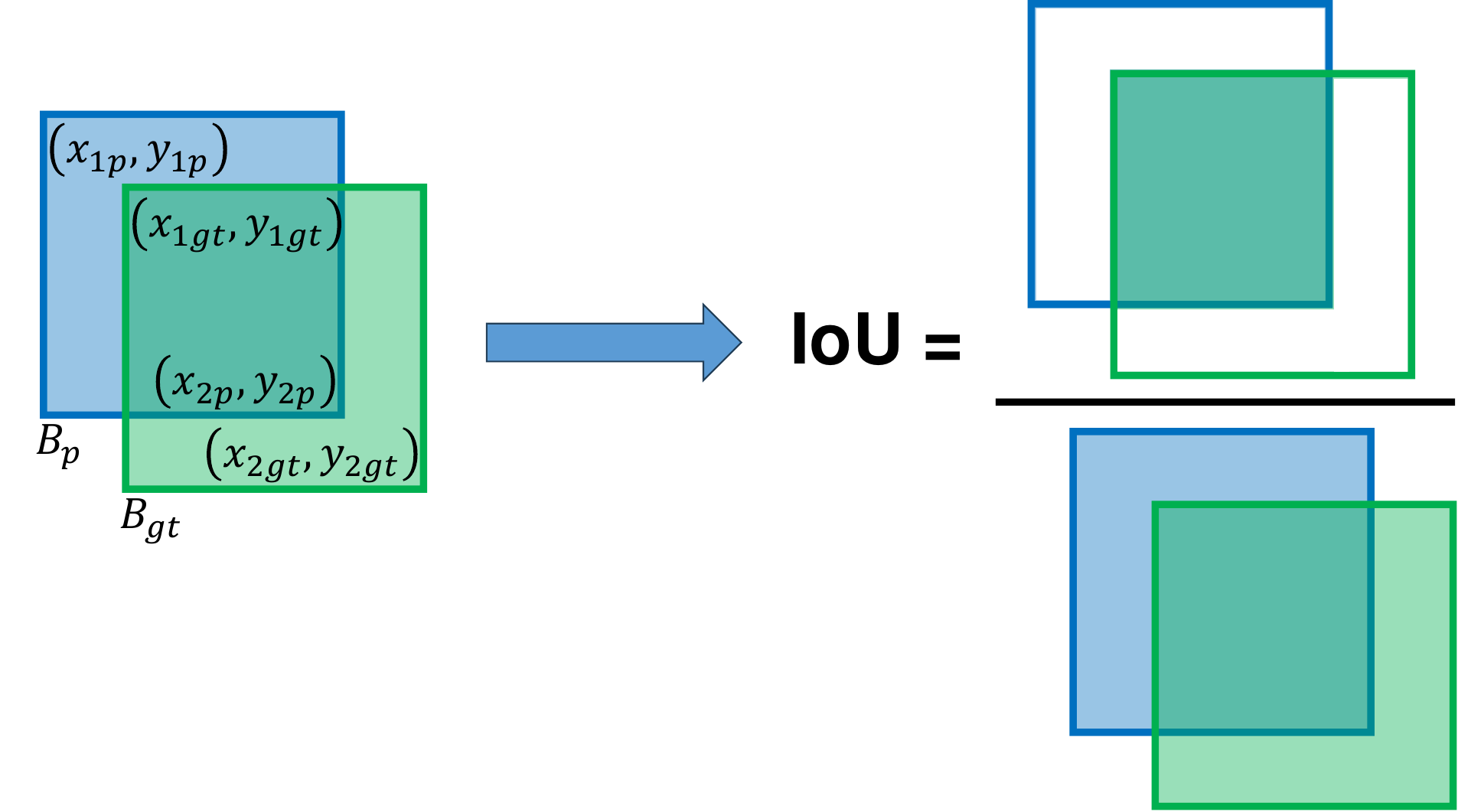}
    \caption{Illustration of the IoU calculation between a detected bounding box $B_p$ and ground truth bounding box $B_{gt}$ with bounding box corners labeled accordingly.}    
    \label{IoUFigure}   
\end{figure}

The intersection-over-union (IoU) is a common metric used in the evaluation of AP and AR to assess object detector performance. The IoU measures how well a predicted bounding box $B_p$ matches with a ground truth bounding box $B_{gt}$ in a given image by quantifying the amount of overlap between $B_p$ and $B_{gt}$, as shown in Figure \ref{IoUFigure}. If we use the notation $\vert B\vert$ to denote the area of a region $B$ in the image, then the IoU is defined as the following:
\begin{equation}
    \label{iouEqn1}
    \text{IoU} = \frac{\vert B_p\cap B_{gt}\vert}{\vert B_p\cup B_{gt}\vert} = \frac{\text{Area of the intersection}}{\text{Area of the union}}.
\end{equation}
Because the intersection $B_p\cap B_{gt}$ is a subset of the union $B_p\cup B_{gt}$, we see that the IoU lies between \num{0} and \num{1}. Thus, the closer the IoU is to \num{1}, the better the match, and the better the performance of the detector. Figure \ref{IoUFigure} depicts the IoU calculation between $B_p$ and $B_{gt}$.

To calculate the IoU, equation \eqref{iouEqn1} can be simplified by using the following identity:
\begin{equation}
\label{areaIntIdentity}
\vert B_p\cup B_{gt} \vert = \vert B_p \vert +\vert B_{gt} \vert-\vert B_p\cap B_{gt}\vert.\end{equation} Substituting \eqref{areaIntIdentity} into \eqref{iouEqn1} leads to the following formula used in this work:
\begin{equation}
\label{iouEqn2}
    \text{IoU} = \frac{\vert B_p\cap B_{gt}\vert}{\vert B_p\vert + \vert B_{gt}\vert-\vert B_p\cap B_{gt}\vert}.
\end{equation}
 In applying \eqref{iouEqn2}, we use the corners of $B_p$ and $B_{gt}$ labeled in Figure \ref{IoUFigure} to calculate the areas $\vert B_p\vert$ and $\vert B_{gt}\vert$ as $\vert B_p\vert=(x_{2p}-x_{1p})(y_{2p}-y_{1p})$ and $\vert B_{gt}\vert=(x_{2gt}-x_{1gt} )(y_{2gt}-y_{1gt})$. For $|B_p\cap B_{gt}\vert$, we use the fact that $B_p\cap B_{gt}$ is either empty, a line segment, or a rectangle. By letting $a_1=\max\{x_{1p},x_{1gt}\}$, $a_2=\min\{x_{2p},x_{2gt}\}$, $b_1=\max\{y_{1p},y_{1gt}\}$ and $b_2=\min\{y_{2p},y_{2gt}\}$, we calculate $\vert B_p\cap B_{gt}\vert$ using the following formula:
\begin{equation}
    \label{areaIntEqn}
    \vert B_p\cap B_{gt}\vert = 
    \begin{cases}
        (a_2-a_1)(b_2-b_1),\quad & a_1\le a_2 \ \text{and} \ b_1\le b_2 \\
        0, \quad & \text{otherwise}
    \end{cases}
    .
\end{equation}

\subsection{Average Precision}
\label{avgPrecision}
Average precision (AP) is the most common metric used to evaluate the accuracy of object detectors. AP is calculated based on the concepts of precision and recall. Precision measures the correctness of the detections made, while recall measures the ability of the detector to correctly detect all ground-truth objects. But before elaborating further, we state the following definitions concerning the detections made by an object detector:
\begin{itemize}
    \item True positive: a correct detection of a ground truth object.
    \item False positive: a detection of an object that is not a ground truth object, or an incorrect detection of a ground truth object.
    \item False negative: an undetected ground truth object.
\end{itemize}
The IoU determines whether a detection is correct or incorrect. A threshold $t$ is chosen and a detection of a ground truth object is considered correct, or a true positive, if $\text{IoU}\ge t$, otherwise it is regarded as a false positive.

Now we define precision and recall as follows using similar notation as \citet{Padilla2020Survey}. Let $G$ denote the total number of ground truth objects in some dataset and let $N_d(\tau)$ be the total number of detections at a given confidence level $\tau$ output by some object detector. Assume that the number of correct detections, i.e., the number of true positives, at the given confidence level is $S(\tau)$. Then precision and recall are defined as:
\begin{equation}
P(\tau) = \frac{S(\tau)}{N_d(\tau)} = \frac{\text{Number of true positives at confidence level } \ \tau}{\text{Total number of detections at confidence level} \ \tau}\label{pEqn}
\end{equation}
\begin{equation}
R(\tau) = \frac{S(\tau)}{G} = \frac{\text{Number of true positives at confidence level} \ \tau}{\text{Total number of ground truth objects}}.\label{rEqn}
\end{equation}

The procedure to calculate $P(\tau)$ and $R(\tau)$ is as follows: set an IoU threshold $t$ for correct detections, fix a confidence threshold $\tau$ and calculate $G$; then retrieve all detections with confidence scores greater than or equal to $\tau$ and sum the number of  detections to obtain $N_d(\tau)$. Finally, determine the number of correct detections $S(\tau)$ at the confidence threshold $\tau$, then apply equations \eqref{pEqn} and \eqref{rEqn}.

From the definitions, we observe the following: first, it is evident that both precision and recall range from \num{0} to \num{1}; and second, recall is a decreasing function of $\tau$ since increasing the confidence threshold decreases $S(\tau)$. This is because $G$, the number of ground truth objects, is independent of the confidence threshold. On the other hand, no obvious mathematical inference can be made about precision because both $S(\tau)$ and $N_d(\tau)$ can change as $\tau$ varies. However, as pointed out by \citet{Padilla2021Comparative}, a good object detector should detect all ground truth objects (no false negatives), while detecting only relevant objects (no false positives). Thus, it is reasonable to conclude that an object detector performs well if its precision stays high as recall increases and confidence decreases. A high area under the precision-recall curve indicates this condition.

\begin{figure}
    \begin{subfigure}{0.5\textwidth}
        \centering  \includegraphics[width=0.9\linewidth]{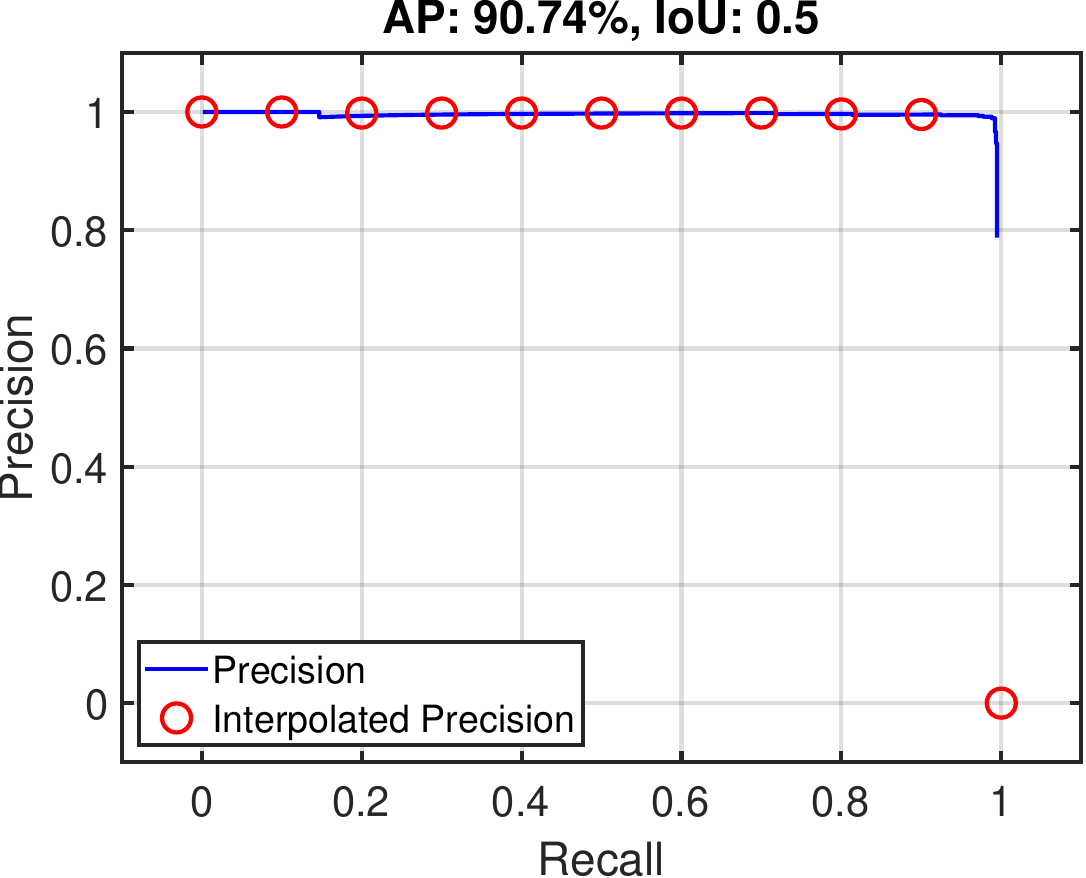}
        \caption{}
        \label{precisionRecall11ptIou50}
    \end{subfigure}%
    \begin{subfigure}{0.5\textwidth}
        \centering
        \includegraphics[width=0.9\linewidth]{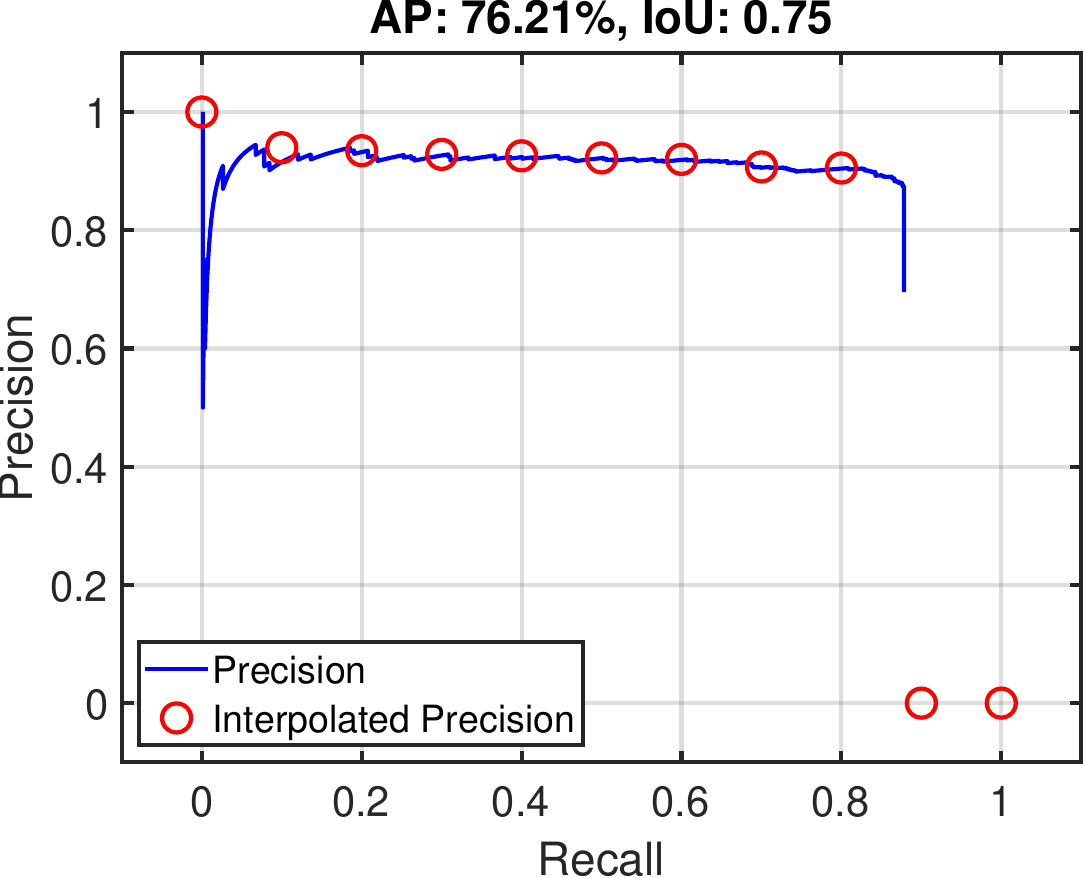}
        \caption{}
        \label{precisionRecall11ptIou75}
    \end{subfigure}
       \caption{Plots of precision against recall from the UAV dataset for two different IoU thresholds using the 11-point interpolation method. The IoU threshold is \num{0.5} in (\subref{precisionRecall11ptIou50}) and \num{0.75} in (\subref{precisionRecall11ptIou75}).}
       \label{precisionRecall11pt}
\end{figure}

AP is a measure of the area under the precision-recall curve. Formally, AP is defined as the average value of the Riemann integral of $P(R)$ on $[0,1]$ in the following:
\begin{equation}
\label{apEqn}
AP = \int_0^1 P(R)\,dR.
\end{equation}
The higher the value of AP, the better the detector can detect ground truth objects while detecting only relevant objects, which means the better the performance of the detector. The integral is evaluated numerically. But before calculating AP, one must first interpolate precision values over a set of reference recall values because $P(\tau)$ is not a function of $R(\tau)$, as can be seen by the zig-zag pattern illustrated in Figure \ref{precisionRecall11ptIou50} and Figure \ref{precisionRecall11ptIou75}. This pattern arises when a false positive is detected because precision will drop while recall stays the same since recall does not depend on false positives. These drops can be quite large, especially near small recall values, as seen in Figure \ref{precisionRecall11ptIou75}. Consequently, two values of precision occur for one value of recall when a false positive is detected. The point of interpolating is to remove this behavior by converting precision into a monotonically decreasing function of recall.

We now outline the procedure to calculate AP similar to \citet{Padilla2020Survey}. First, given a dataset $D$ of detections and ground truths, set an IoU threshold $t$ and order the detections output by the detector by increasing confidence score. Then apply equations \eqref{pEqn} and \eqref{rEqn} to calculate the set $\{R(\tau),P(\tau)\}$ of precision and recall pairs at each confidence level $\tau$ in $D$. 

From $D$, choose an ordered set $\{\tau(k)\}$ of $K$ confidence values output by the object detector such that:
\begin{equation}
\label{confidenceCondition}
 \tau(k)<\tau(k+1) \quad \text{for} \quad  1\le k\le K-1.
\end{equation}
Then sample a set $\{R(\tau(k)),P(\tau(k))\}$ of $K$ precision and recall values such that:
\begin{equation}
\label{recallCondition}
    R(\tau(k))>R(\tau(k+1)) \quad 1\le k\le K-1.
\end{equation}
Next, set $\tau(0)=0$ and $\tau(K+1)=1$ and add to $\{R(\tau(k)),P(\tau(k))\}$ the following two precision and recall pairs:
\begin{equation}
    \left\{R(\tau(0)),P(\tau(0))\right\} = \left\{1,0\right\}\label{rptau0},
\end{equation}
\begin{equation}
    \left\{R(\tau(K+1)),P(\tau(K+1))\right\} = \left\{0,1\right\}\label{rptau1}.
\end{equation}

After constructing $\{R(\tau(k)),P(\tau(k))\}$, the next step is to apply the continuous function $P_{\text{interp}}(R)$ mapping $[0,1]$ into $[0,1]$ to interpolate precision values at distinct recall points:
\begin{equation}
\label{pInterpEqn}
P_{\text{interp}}(R)=\max_{k|R(\tau(k))\ge R}\{P(\tau(k))\}.
\end{equation}

Next, AP is calculated by numerically evaluating the integral in equation \eqref{apEqn}. There are two approaches to this calculation. Both methods require defining a set of reference recall values and applying equation \eqref{pInterpEqn} to interpolate the precision at these points; the difference is in how these recall values are generated. The two methods are N-point and all-point interpolation.

\subsubsection*{\textit{N-Point Interpolation}}
In this approach, one constructs a set of $N$ equally spaced reference recall values in the interval $[0,1]$ by defining:
\begin{equation}
    \label{rRefNpt}
    R_r (k)=\frac{N-k}{N-1}, \quad 1\le k\le N.
\end{equation}
Using these recall values, the precision values are interpolated by sampling $P_{\text{interp}}(R)$ in equation \eqref{pInterpEqn} at each point in $\{R_r(k)\}$. AP is then calculated as follows:
\begin{equation}
\label{apNpt}
AP=\frac{1}{N}\sum_{k=1}^N P_{\text{interp}}(R_r(k)).
\end{equation}
Popular choices for calculating AP using this method are $N=11$ and $N=101$.

\subsubsection*{\textit{All-Point Interpolation}}
For the all-point interpolation method, the reference recall values are those constructed from equations \eqref{confidenceCondition}--\eqref{rptau1}, which we restate here:
\begin{equation*}
R_r(k) = R(\tau(k)), \quad  0\le k\le K+1.
\end{equation*}
Then AP is evaluated as:
\begin{equation}
\label{apAllpt}
AP=\sum_{k=0}^K(R_r(k)-R_r(k+1))P_{\text{interp}}(R_r(k)).
\end{equation}
This is the standard approach since the \num{2014} Pascal VOC object detection challenge \citep{Everingham2014Pascal}.

No matter which approach is applied to calculate AP, it is dependent on the IoU threshold $t$. Thus, this value of $t$ must be reported in addition to AP. Because of this dependence on the IoU threshold, there are several variants of AP used to evaluate object detection algorithms. We leverage the most popular  \citep{Everingham2014Pascal,Redmon2016You,Ren2017Faster,Padilla2020Survey,Padilla2021Comparative} in this work:
\begin{itemize}
	\item AP50: $t=0.5$.
	\item AP75: $t=0.75$.
	\item AP@50:5:95: the average of AP from $t=0.5$ to $t=0.95$ in $0.05$ increments.
 \end{itemize}

\subsection{Average Recall}
\label{avgRecall}
Average recall (AR) is another metric that measures the localization ability of an object detector. AR rewards for high recall and localization accuracy of the detector and has been shown to correlate well with AP \citep{Hosang2016What}. To calculate AR, one first defines a function $R_{\text{IoU}}(t)$ mapping an IoU threshold $t$ in $[0.5,1]$ to a recall value $R$ in $[0,1]$. Then, AR is defined as the average value of $R_{\text{IoU}}(t)$ on $[0.5,1]$ in the following:
\begin{equation}
\label{arEqn}
 AR=2\int_{0.5}^{1} R_{\text{IoU}}(t)\,dt.   
\end{equation}

As with AP, the Riemann integral in equation \eqref{arEqn} is evaluated numerically. \citet{Hosang2016What} propose a simple formula that approximates the integral over a discrete set of $G$ ground truth objects. The formula is:
\begin{equation}
\label{arNumEqn}
AR = \frac{2}{G}\sum_{i=1}^G\max(\text{IoU}_i-0.5,0),
\end{equation}
where $\text{IoU}_i$ is the largest IoU between a predicted bounding box and ground-truth $i$.

The procedure for evaluating AR is straightforward. For each ground truth object in some dataset $D$, use equations \eqref{iouEqn2} and \eqref{areaIntEqn} to calculate the IoU between the given ground truth box and each associated predicted bounding box; then take the largest IoU and calculate $\max(\text{IoU}_i-0.5,0)$. After obtaining all such values for each ground truth object, apply equation \eqref{arNumEqn}.

The main limitation of AR is that, as the number of detection proposals made by an object detector increases, the value of AR approaches \num{1}. Thus, in practice, when calculating AR, one limits the number of detection proposals for each ground truth object. This leads to different metrics as the value of AR can vary when the number of detection proposals changes. The two most common variants, which are evaluated in this work, are:
\begin{itemize}
    \item AR1: The number of detection proposals is limited to \num{1}.
    \item AR10: The number of detection proposals is limited to \num{10}.
\end{itemize}

\section{System Model}
\label{gunTurretSystem}
\subsection{Turret Model}

\begin{figure}
    \centering
    \includegraphics[scale=0.5]{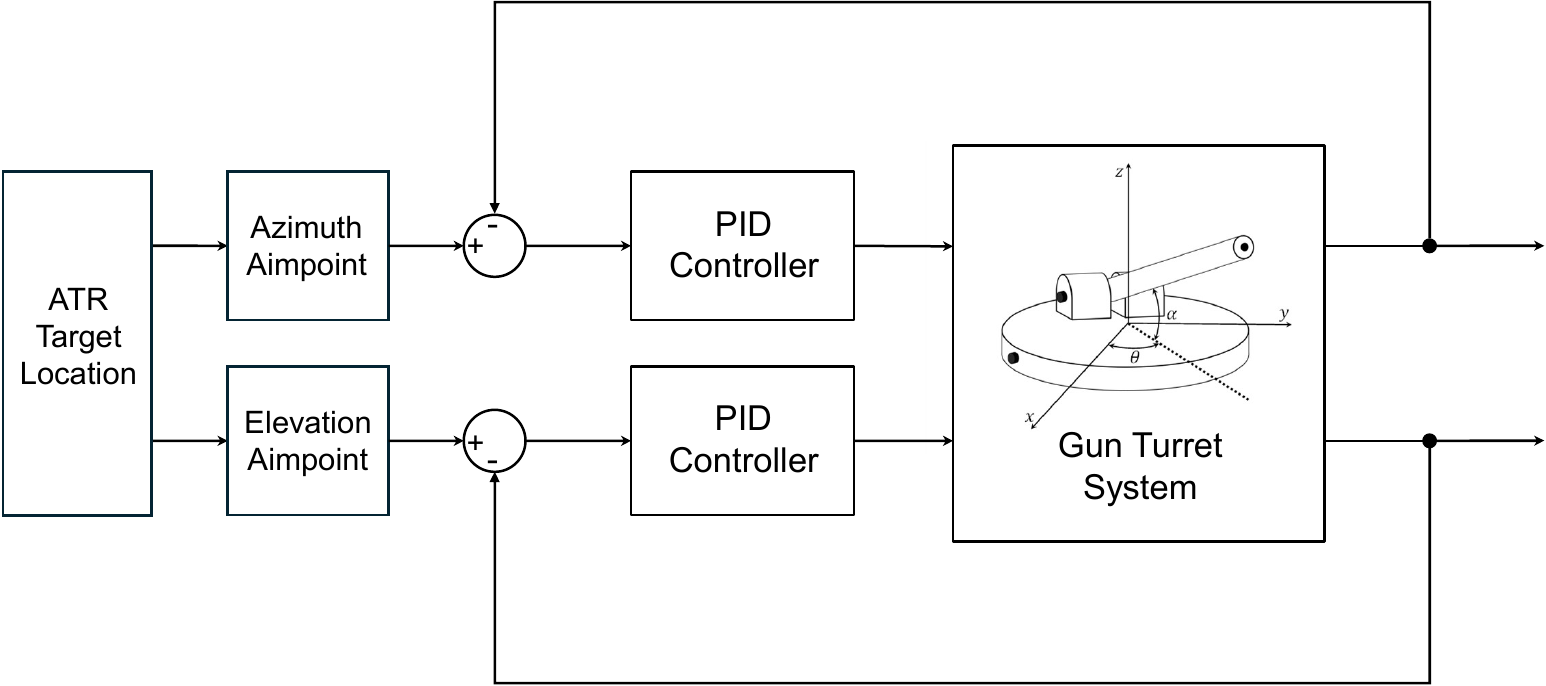}
    \caption{Block diagram of the control system.}
    \label{gunTurretBlockDiagram}
\end{figure}

 The gun turret model used in our controlled targeting simulations is comprised of two rigid bodies: a platform, modeled as a uniform disk, and a gun barrel, modeled as a uniform rod. An illustration of the gun turret with the outputs labeled is shown in Figure \ref{gunTurretBlockDiagram}. The inertial frame $\{x,y,z\}$ is fixed to the ground with the origin at the center of mass of the platform. The inputs to the system are torques supplied by two DC motors which drive the platform and the gun barrel. We denote these inputs as $u_1$ for the platform and $u_2$ for the gun barrel, each measured in the SI unit of torque \si{\newton\cdot\meter}. Referring to Figure \ref{gunTurretBlockDiagram}, the outputs are $\theta$, the azimuth position of the platform measured from the fixed $x$-axis, and $\alpha$, the elevation position of the gun barrel measured from the dashed line that rotates with the platform; the outputs are measured in \si{\radian}. 
 
 Focusing on stationary target scenarios, we assume that the rotational speeds of each rigid body are limited to \SI{45}{\degree/\second} and the elevation is limited to \SI{30}{\degree}. Additionally, we assume that the pivot point of the gun barrel coincides with the center of mass of the platform. These assumptions allow us to ignore coupling effects due to Coriolis and centrifugal forces which are proportional to products of the angular speeds. We also assume the gun barrel starts moving from a position of static equilibrium and that the inertia of each DC motor is negligible. Then, using the free-body diagram of the system shown in the Appendix, we sum the moments acting on each body and apply Newton’s second law to obtain the following two equations of motion:
\begin{align}
J_1\ddot{\theta} + b_1\dot{\theta} &= u_1	\label{thetaeqnofmotion}\\
J_2\ddot{\alpha} + b_2\dot{\alpha} &= u_2, \label{alphaeqnofmotion}
\end{align}
where
$J_1 = \frac{1}{2}m_1 R^2+\frac{1}{3} m_2 L^2$ and $J_2 = \frac{1}{3} m_2 L^2$.
Equation \eqref{thetaeqnofmotion} describes the dynamics of the platform, while equation \eqref{alphaeqnofmotion} describes the dynamics of the gun barrel. Both equations include damping due to friction at the bearings. Note that the assumptions on the elevation limit of \SI{30}{\degree} and gun barrel movement from static equilibrium allow us to drop the gravitational torque term $\frac{1}{2}m_2gL\cos\alpha$ that appears in equation \eqref{alphaeqnofmotionderiv} in the Appendix. Additionally, the total moment of inertia of the platform $J_1$ is estimated from $J(\alpha) = \frac{1}{2}m_1R^2+\frac{1}{3}m_2L^2\cos^2\alpha$ by applying the elevation limit assumption; this formula assumes the pivot point of the gun barrel coincides with the center of the platform. The definition of the model parameters and their values are shown in Table \ref{lineargunturretparametervalues}. For purposes of this study, these values are chosen based on consideration of previous work on turret control systems \citep{carlstedt2021modelling,idris-hudha-ASCC2015,lyth2021modelling,Ma2022adaptive,Nasyir2014,rahmat-et-al-AIMT2016,Xia2016modeling,Yuan2021precision,Yuan2024bidirectional}, but are not specific to a vehicle.

Assuming both the azimuth and elevation start from \SI{0}{\degree}, by using Laplace transforms, we convert equations \eqref{thetaeqnofmotion} and \eqref{alphaeqnofmotion} from the time domain to an equivalent system of equations in the frequency domain, denoting $s$ as the frequency variable with units of \si{\radian/\second}. The advantage of this procedure is that we can analyze the system algebraically and derive the following two transfer functions, one between the input $u_1$ and the output $\theta$, and the other between the input $u_2$ and the output $\alpha$, for controller design:
\begin{align}
G_1 (s) &=\frac{1}{J_1 s(s+c_1 )}, \label{thetaTransferFunction} \\
G_2 (s) &=\frac{1}{J_2 s(s+c_2 )},\label{alphaTransferFunction}
\end{align}
where $c_1 = \frac{b_1}{J_1}$ and $c_2 = \frac{b_2}{J_2}$.

\begin{table} 
    \centering
    \caption{Parameters of the gun turret model}
    \begin{tabular}{cll}
        \hline
        \textbf{Parameter} & \multicolumn{1}{c}{\textbf{Description}} & \textbf{Value} \\
         \hline
         $m_1$ & Mass of the platform & \num{8.67e3} [\si{\kilogram}] \\
         $m_2$ & Mass of the gun barrel & \num{4.97e3} [\si{\kilogram}] \\
         $b_1$ & Damping coefficient of the platform & \num{6.00e4} [\si{\newton\cdot\meter/\second}] \\
         $b_2$ & Damping coefficient of the gun barrel & \num{6.00e4} [\si{\newton\cdot\meter/\second}] \\
         $J_1$ & Moment of inertia of the platform & \num{7.99e4} [\si{\kilogram\cdot\meter^2}] \\
         $J_2$ & Moment of inertia of the gun barrel & \num{4.83e4} [\si{\kilogram\cdot\meter^2}] \\
         $R$ & Radius of the platform & \num{2.70} [\si{\meter}] \\
         $L$ & Length of the gun barrel & \num{5.40} [\si{\meter}] \\
         \hline
    \end{tabular}
    \label{lineargunturretparametervalues}
\end{table}

\subsection{PID Control}
PID control is a popular method of process control used throughout industry for its simple design and effectiveness in control tasks such as output reference tracking. For this task, we define the error between the reference $r(t)$ and the measured output $y(t)$ as $e(t)=r(t)-y(t)$. The PID control law in the time domain is then defined in the following equation:
\begin{equation}
\label{pidControlEqn}
u(t)= K_P e(t) + K_I\int e(t)\, dt + K_D\dot{e}(t),
\end{equation}
where $K_P$, $K_I$ and $K_D$ are the proportional, integral and derivative gains to be set by the designer. 

For this work, we use frequency response methods with equations \eqref{thetaTransferFunction} and \eqref{alphaTransferFunction} to design two PI+lead controllers to simulate controlled aiming of the gun turret, one for the azimuth and one for the elevation. This type of controller approximates pure PID control. The integral term improves steady state accuracy, while the lead, which approximates derivative control, can speed up the response \citep{FranklinCh6FreqResp}. We design two controllers because the small speed assumption decouples the outputs of the turret system and enables us to consider the two inputs separately. The transfer function of a PI+lead controller takes on the following form in the frequency domain:
\begin{equation}
\label{piLeadControlEqn}
C(s)= K_P \left(\frac{s+1/T_I}{s}\right)\left(\frac{T_D s+1}{\gamma T_D s+1}\right),
\end{equation}
where $0<\gamma<1$, $T_D>0$ and $T_I>0$. 

In the design procedure, we specify a bandwidth requirement for speed and a phase margin (PM) requirement for stability. The bandwidth $\omega_{BW}$ is an estimate of the speed of the closed-loop response. We use the gain-crossover frequency $\omega_{gc}$, which is the frequency at which the magnitude of the loop gain $L(j\omega)=G(j\omega)C(j\omega)$ is unity, to ensure the bandwidth requirement is met by the following rule of thumb $\omega_{gc}\le \omega_{BW}\le 2\omega_{gc}$. The PM measures the degree to which stability conditions are met, which are requirements on the magnitude and phase of $L(j\omega)$. The requirement for stability is $\vert L(j\omega)\vert < 1$ at $\angle G(j\omega)=-180\si{\degree}$ for systems where increasing the controller gain leads to instability \citep{FranklinCh6FreqResp}. The design procedure for our controllers is described in the Appendix and a block diagram of the control system is shown in Figure \ref{gunTurretBlockDiagram}.

The control objectives are to stabilize the system and to quickly move the gun turret to the desired aimpoint. We use the settling time $t_s$ to assess the design objectives, which we define as the time for the tracking error to reach and remain within \SI{0.1}{\percent} of the aimpoint. We tune the controllers to achieve a $1$ second settling time. Assuming static targets, this means the horizontal error is at most \SI{0.1}{\percent} of the target’s horizontal aimpoint and the vertical error is at most \SI{0.1}{\percent} of the target’s vertical aimpoint $1$ second after the gun barrel starts to move. This choice is based on an empirical result for accuracy in stationary target scenarios wherein a \SI{1}{mil} aiming error corresponds to \SI{1}{\meter} miss distance at \SI{1000}{\meter} range \citep{strohm2013introduction}. Note that we use the \si{mil} as defined by the North Atlantic Organization Treaty (NATO), i.e., an angular measurement equal to $1/6400$ of a circle, or $\si{360\degree}/6400 = 0.05625\si{\degree}$  \citep{weaver1990system}. The parameters of our controllers designed in this study are shown in Table \ref{azimuthelevationcontrollerparameters}. After initial tuning of our controllers, we have increased the proportional gain $K_P$ in both by \SI{50}{\percent} to meet the settling time objective; the reported value of $K_P$ reflects this increase. The observed settling time for both of our controllers is \SI{1.0}{\second}.

\begin{table}
    \centering
     \caption{PID controller parameters for the gun turret system. The last two columns report the measured $\omega_{gc}$ and PM for each controller after the design procedure.}
    \begin{tabular}{cccccccc}
    \hline
    \textbf{Variable} & \textbf{Controller} & $K_P$ & $T_D$ [\si{\second}] & $T_I$ [\si{\second}] & $\gamma$ &  $\omega_{gc}$ [\si{\hertz}] & $\text{PM}$ [\si{\degree}] \\
    \hline
     $\theta$ & PI+lead$_{\theta}$ & \num{3.33e7} & \num{0.17} & \num{0.22} & \num{0.017} & \num{10.8} & \num{70.7} \\
    $\alpha$ & PI+lead$_{\alpha}$ & \num{4.48e6} & \num{0.32} & \num{0.50} & \num{0.024} & \num{4.70} & \num{69.7}
    \\
    \hline
    \end{tabular}
    \label{azimuthelevationcontrollerparameters}
\end{table}

\section{Evaluating the Probability of a Hit}
\label{evaluatePh}
We follow a similar procedure as in the Tank Accuracy Model (TAM) report by \citet{bunn1993tank} to calculate the probability of a hit in our simulations. A hit is regarded as a hit in the horizontal coordinate, denoted $X$, and a hit in the vertical coordinate, denoted $Y$. These random variables are assumed to be continuous, independent and arising from normal distributions in each respective direction. The sources of error considered in TAM consist of biases and random errors in the $X$ and $Y$ directions that are assumed to be mostly uncorrelated and normally distributed. The biases arise from effects such as weapon cant, variation in muzzle velocity and crosswind, for example \citep{bunn1993tank,strohm2013introduction}. We combine the horizontal biases together forming the mean $\mu_x$ of the impact distribution in the $X$ direction; we do the same for the vertical biases, denoting $\mu_y$ as the mean of the impact distribution in the $Y$ direction. We follow the same procedure with the variances of all the random error sources, which include the ammunition dispersion and the random component of the turret movement error. We denote $\sigma_x^2$ as the total variance of all the random errors in the $X$ direction and $\sigma_y^2$ as the total variance of all the random errors in the $Y$ direction. The probability of a hit is then evaluated as:
\begin{equation}
\label{phEqn}
P_h=\int_{x_1}^{x_2}\frac{1}{\sqrt{2\pi}\sigma_x} e^{-\frac{(x-\mu_x)^2}{2\sigma_x^2}}\,dx \int_{y_1}^{y_2}\frac{1}{\sqrt{2\pi}\sigma_y} e^{-\frac{(y-\mu_y )^2}{2\sigma_y^2}}\, dy,
\end{equation}
where $x_1$ and $x_2$ are the horizontal limits of the target and $y_1$ and $y_2$ are the vertical limits of the target. The limits of integration in equation \eqref{phEqn} are acquired from the ground truth data used to train the AI object detection algorithm. To calculate the integrals, we use the \texttt{normcdf} function in MATLAB \citep{MatlabVersion} which approximates the cumulative distribution function of the standard normal distribution with zero mean and unit variance. 
\section{Results}
\label{resultsanddiscussion}

In this section, we discuss the results of our assessment on the performance of our controlled gun turret system given target location input from an object detection model. The object detection model has been trained with an open source Faster R-CNN architecture \citep{Ren2017Faster} on a set of \num{760} RGB images of an unmanned aerial vehicle (UAV). The output dataset consists of \num{955} bounding box coordinates and confidence score predictions made by the object detector evaluating the set of UAV imagery used for training the model. The format of the bounding box data is $[x_1,y_1,x_2,y_2 ]$ where $(x_1,y_1)$ and $(x_2,y_2)$ are the top-left and bottom-right corners of the box in pixels (note that $y_1<y_2$ in pixel coordinates). We have performed our data processing and simulations in MATLAB \citep{MatlabVersion}. 

In our analysis, we generate a dataset of the AI error defined in Section \ref{metricsForObjectDetectors} to determine if this error significantly correlates with the following metrics: the confidence score, IoU and area of the detected bounding box. We then simulate controlled targeting with the gun turret given target location input from the object detection model in two different starting points for aiming -- the bottom left corner of the image and the center of the image. The simulation data is used to analyze the impacts on the errors in moving the turret to the aimpoint. For this analysis, we compare error distribution histograms and statistics between the two different targeting scenarios. These results are discussed in Section \ref{threeErrorDistributions}. We then conduct additional simulations with our controlled gun turret system to determine which object detection performance metrics correlate with the probability of a hit. These simulations are done using five smaller datasets sampled from the \num{955} detections that make up the entire output dataset of the object detector from evaluating the UAV imagery. The metrics we have evaluated for this analysis are the confidence score, IoU and the variants of AP and AR described in Section \ref{metricsForObjectDetectors}. These results are summarized in Section \ref{metricsAndPh}.

In each simulation, the aimpoints are estimated from the detected bounding box centroid data using the formulas $x_b/\rho$ for the azimuth and $y_b/\rho$ for the elevation, where $\rho$ is the target range. As the images have been acquired in a manner where the range is unknown, and given target size is considered somewhat independent of range due to magnification, we assume a fixed range to obtain the azimuth and elevation aimpoints. Using these angles, we calculate two reference commands for each gun turret output to track as the turret moves to the aimpoint. These commands are modeled as ‘move-and-settle’ inputs, which are a combination of a ramp input and a constant input after the gun turret is near the aimpoint. The general form of the reference command is given in the following equation:
\begin{equation}
\label{rampinput}
    r(t) = \begin{cases}
                0, & t < 0,  \\
                Vt, & 0 \le t \le P/V \\
                P, & t > P/V
            \end{cases}
            ,
\end{equation}
where $V$ is the slope of the ramp and $P$ is the aimpoint. Note $P = x_b/\rho$ for the azimuth and $P = y_b/\rho$ for the elevation. For both reference commands, we set $V$ to \SI{40}{\degree/\second} to limit the slew rate of the platform and gun barrel. This choice is based on work related to modeling electromechanical motors for main battle tanks \citep{carlstedt2021modelling,lyth2021modelling}. The firing time $t_f$ is defined as the amount of time elapsed in a simulation of controlled gun turret movement from stationary position, and is chosen to occur $0.5$ seconds after the move portion in either the azimuth or elevation reference command, whichever takes longer. Furthermore, we state additional definitions of quantities measured in our analysis of AI data and simulation data using the notation introduced in Section \ref{metricsForObjectDetectors}. At the firing time in our simulations, we denote $\mathbf{x}_r$ as the shot. The controller error vector is then defined as $\mathbf{R}_{br} = \mathbf{x}_b-\mathbf{x}_r$ and the total error vector is defined as $\mathbf{R}_{rg} = \mathbf{x}_r-\mathbf{x}_g$. We define $\Vert\mathbf{R}_{br}\Vert$ and $\Vert\mathbf{R}_{rg}\Vert$ as the controller error and the total error, respectively.

\begin{figure}
    \begin{subfigure}{0.5\textwidth}
        \centering
        \includegraphics[width=0.8\linewidth]{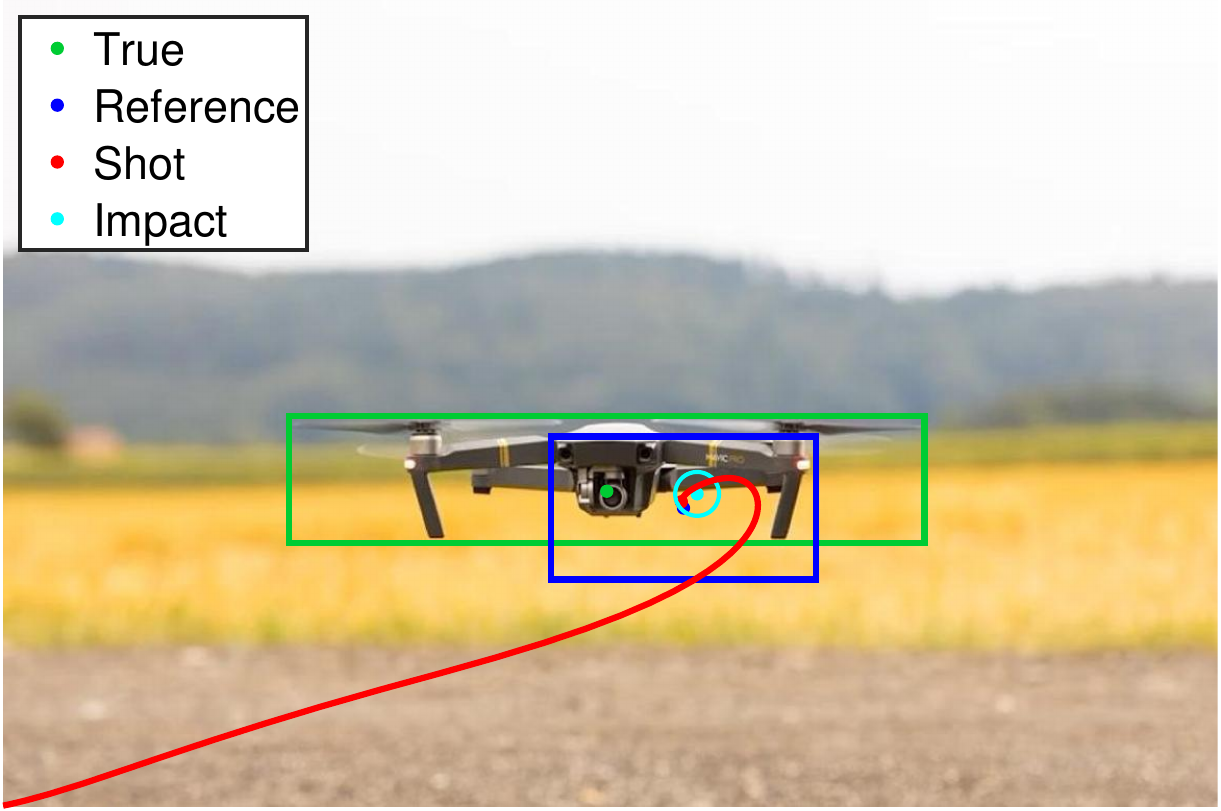}
    \caption{}
    \label{droneImageBtlc}
    \end{subfigure}%
        \begin{subfigure}{0.5\textwidth}
        \centering
        \includegraphics[width=0.8\linewidth]{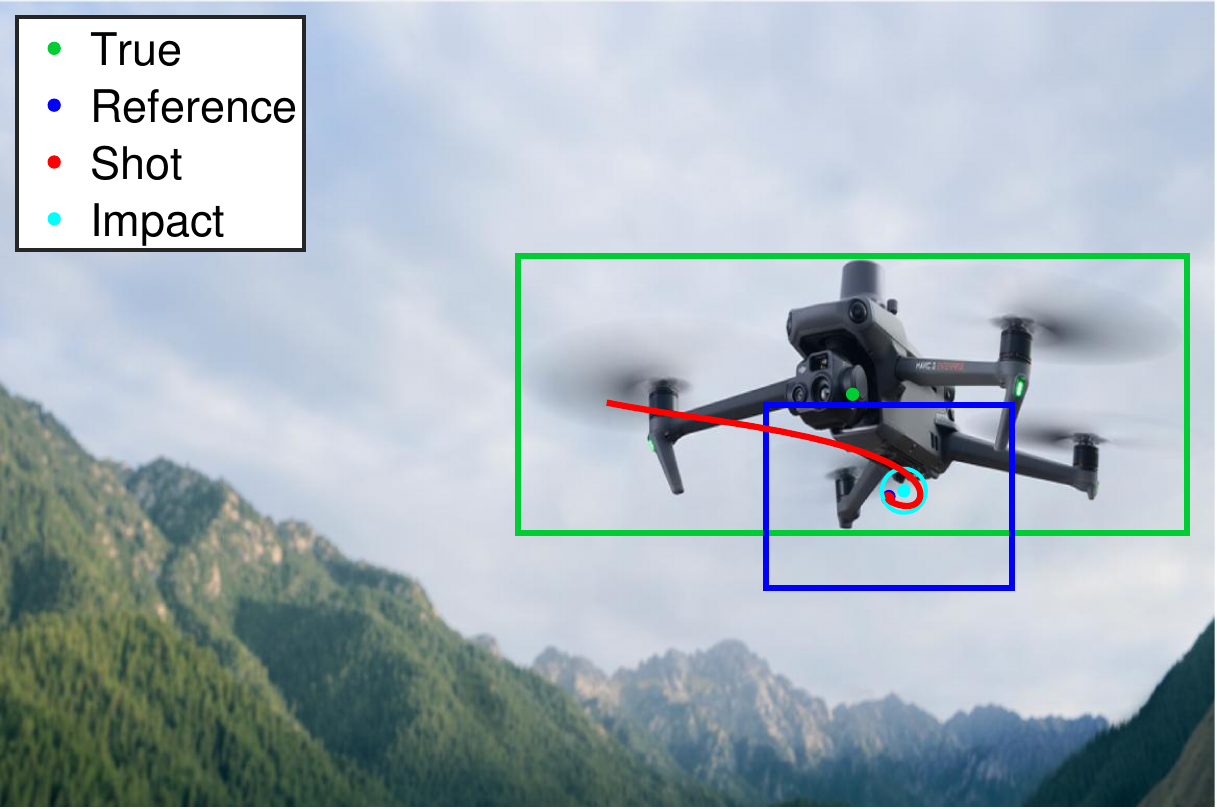}
    \caption{}
    \label{droneImageCenter}
    \end{subfigure}
        \caption{Example output from two targeting simulations. Note, the images are not part of the dataset used to train the object detection model, but are publicly available. In (\subref{droneImageBtlc}), targeting starts from the bottom left corner of the image and in (\subref{droneImageCenter}), targeting starts from the center of the image.}
        \label{droneImage}
\end{figure}

As an example, Figure \ref{droneImage} shows output from two targeting simulations. Both plots show the ground truth bounding box (gray border), detected bounding box (dark gray border), gun barrel path (gray curve) and the impact distribution (light gray ellipse). The circle marker is the centroid of the ground truth bounding box and the asterisk marker is the centroid of the detected bounding box. The square marker is the mean position of the shot at the firing time and the triangle marker is the mean of the impact distribution. The impact distribution, indicated by the light gray ellipse, represents the \SI{98}{\percent} confidence bound for the impact. This means that \SI{98}{\percent} of the observations lie within the bounds of the ellipse. The impact distribution is offset from the shot because we incorporate the additional error sources from the projectile trajectory that are estimated from tabulated values and equations used to calculate the probability of a hit in the TAM report by \citet{bunn1993tank}.

In our results, we discuss statistics such as the mean, standard deviation and correlation coefficient between two variables of different datasets. The mean is calculated as $\mu = \frac{1}{N}\sum_{i=1}^Nx_i$  and the standard deviation is calculated as $\sigma = \sqrt{\frac{1}{N}\sum_{i=1}^N(x_i-\mu)^2}$, where $x_i$ is an observation of a variable and $N$ is the sample size. The correlation coefficient between two variables $x$ and $y$ is calculated as $r_{xy}=\text{Cov}(x,y)/\sigma_x \sigma_y$, where $\text{Cov}(x,y) = \frac{1}{N}\sum_{i=1}^N(x_i-\mu_x)(y_i-\mu_y)$.

Additionally, we state the following assumptions used in this work. We assume all targets are stationary and at a range of \SI{1000}{\meter}, and that each image is calibrated so that \num{34} pixels in the image represents \SI{1}{\meter}. This has been estimated from the images using reference objects. We also model the update of target location by the object detector from processing images in real time to estimate the uncertainty added to the error in moving the controlled turret system to the aimpoint. We assume this procedure occurs at a rate of \SI{0.5}{\hertz} and that the white noise present in each image is normally distributed with \num{0} mean and \num{0.5} standard deviation in units of NATO mils \citep{weaver1990system}. This model and the assumptions on the update procedure and white noise in the image enable us to quantify the uncertainty added to the feedback system using the standard deviation of the turret movement error in \eqref{errorStdevWhiteNoise} with the sample frequency set to $f_s =$ \SI{200}{\hertz}. The model of the target location update and standard deviation calculation are described in Section \ref{errorstdevderivation} of the Appendix.

\subsection{Relationship between Object Detection Metrics and AI Error}
\label{metricsAndAiError}
\begin{figure}
    \begin{subfigure}{0.24\textwidth}
        \centering  \includegraphics[width=\linewidth]{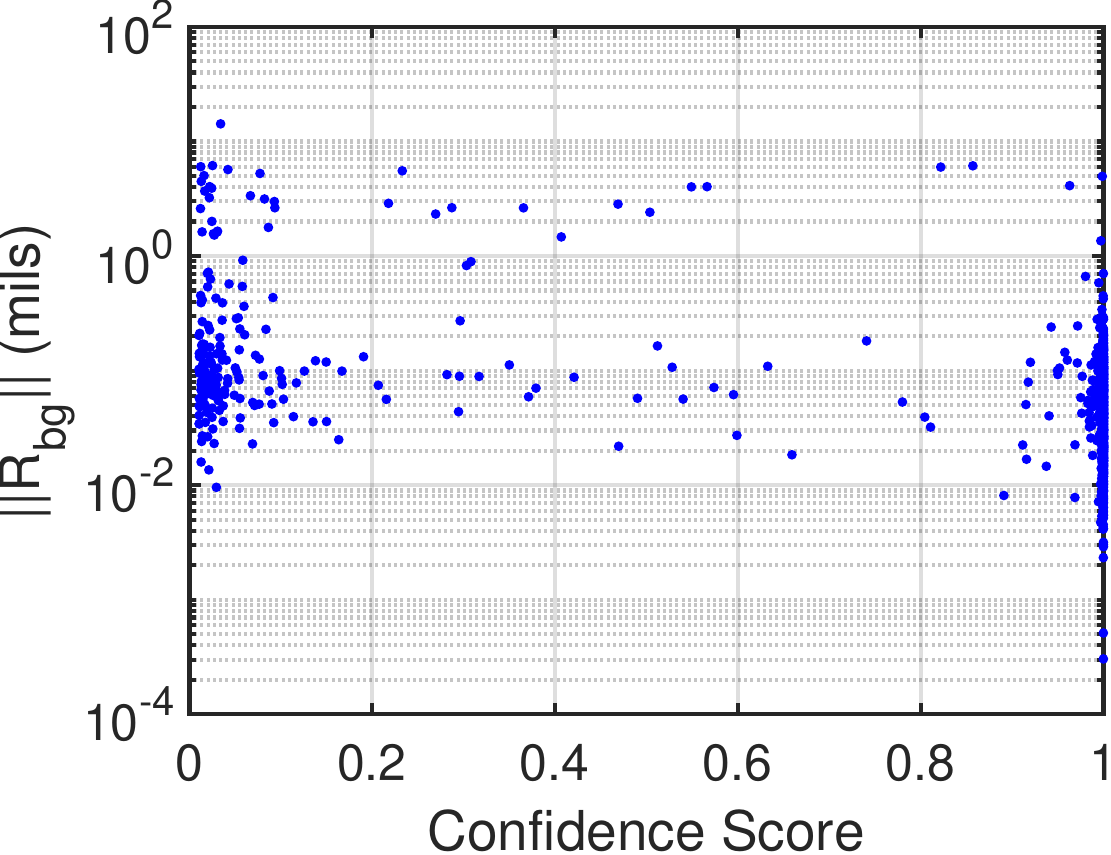}
        \caption{}
        \label{rbgvscscorer1000m}
    \end{subfigure}
    \begin{subfigure}{0.24\textwidth}
        \centering
        \includegraphics[width=\linewidth]{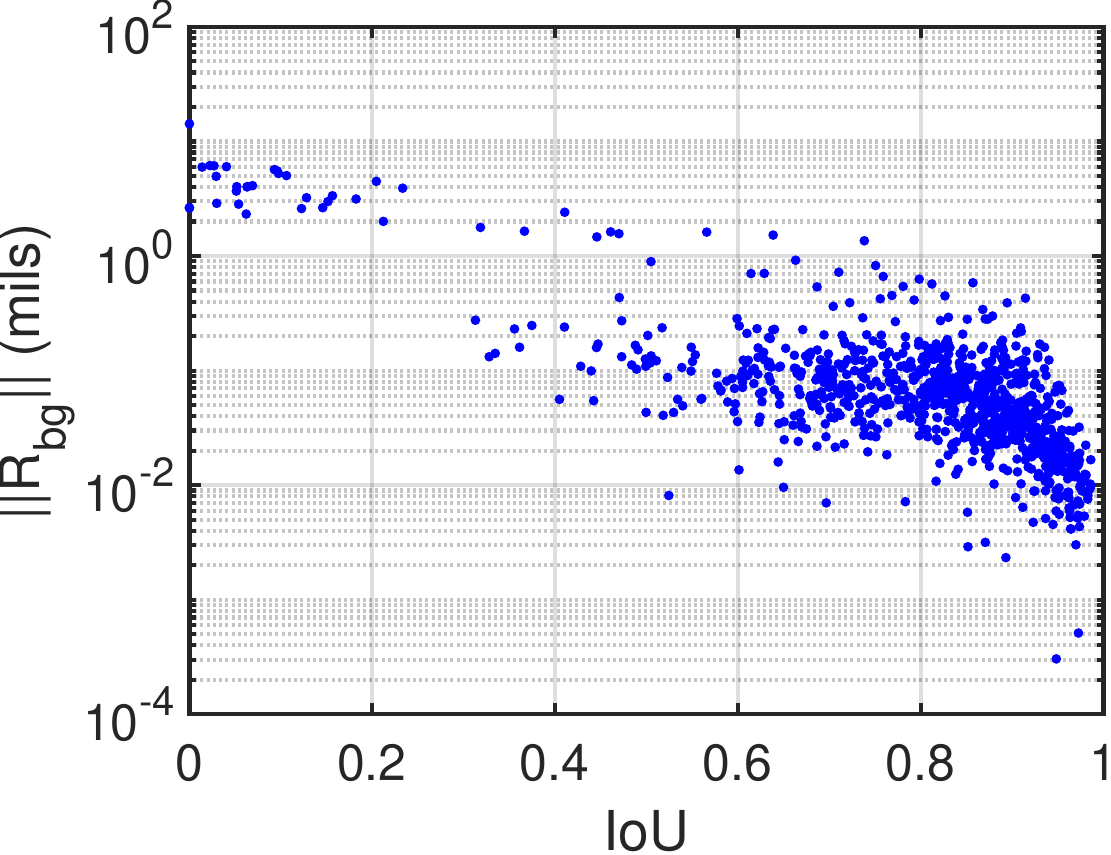}
        \caption{}
        \label{rbgvsiour1000m}
    \end{subfigure}
    \begin{subfigure}{0.24\textwidth}
        \centering
        \includegraphics[width=\linewidth]{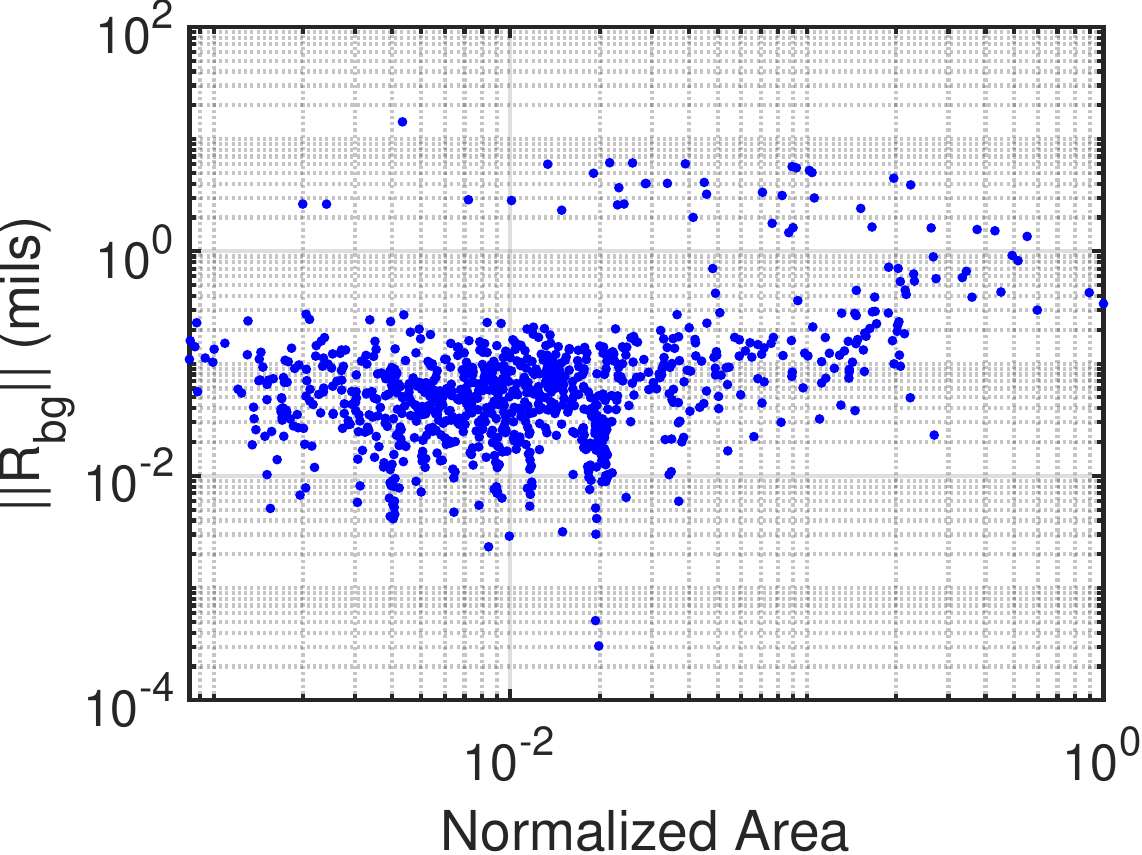}
        \caption{}
        \label{rbgvsarear1000m}
    \end{subfigure}
    \begin{subfigure}{0.24\textwidth}
        \centering
        \includegraphics[width=\linewidth]{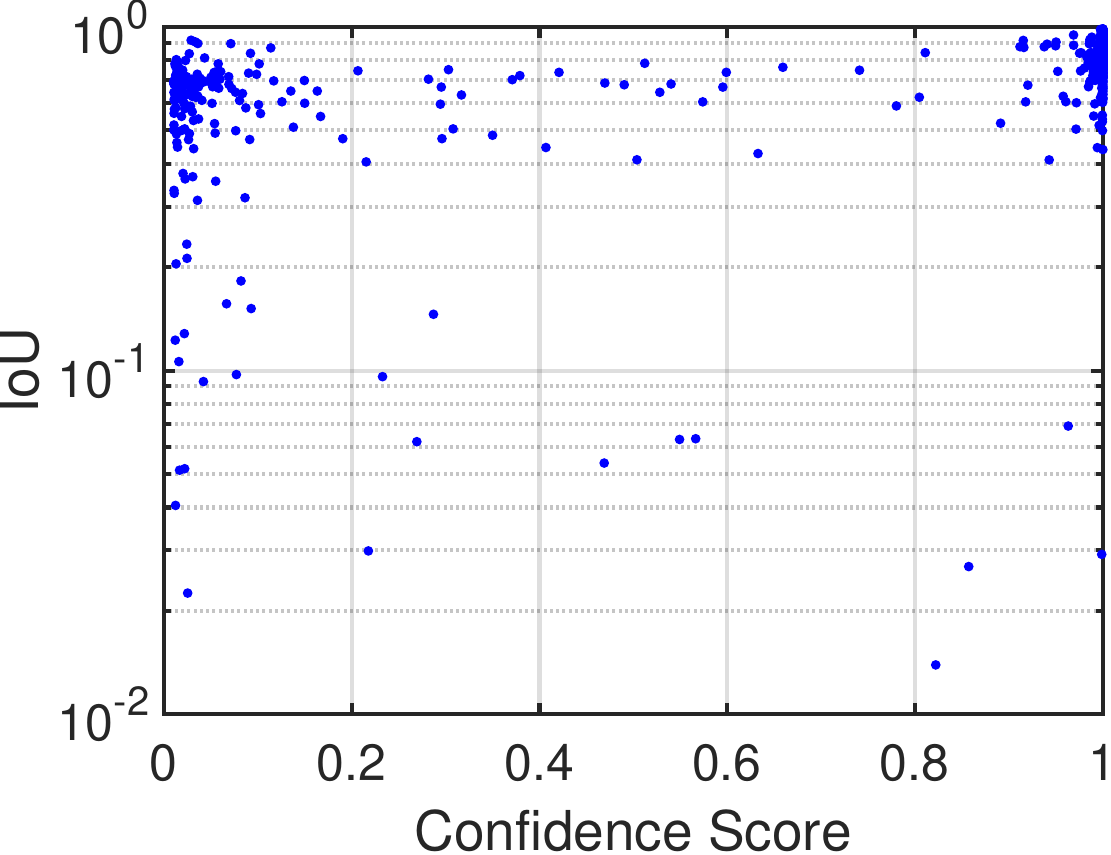}
        \caption{}
        \label{iouvscscorer1000m}
    \end{subfigure}
       \caption{Plots of the AI error against (\subref{rbgvscscorer1000m}) the confidence score, (\subref{rbgvsiour1000m}) the IoU and (\subref{rbgvsarear1000m}) the normalized detected box area; the area is scaled by the largest area out of all detected boxes. Plot (\subref{iouvscscorer1000m}) shows the IoU against the confidence score. The total number of observations in each plot is $N=955$.}
\end{figure}

In this analysis, we examine the AI error $\Vert \mathbf{R}_{bg}\Vert$ as a function of three variables associated with the prediction of the object detection algorithm, namely, the confidence score, IoU and predicted bounding box area.
Beginning with Figure \ref{rbgvscscorer1000m}, we observe no significant linear correlation between the AI error and the confidence score; the correlation coefficient for the entire dataset is \num{-0.29}. Restricting the data to those observations with an IoU of \num{0.5} and above or confidence scores of \num{0.99} and above does not reveal any significant correlation either. The correlation coefficient for samples with an IoU of \num{0.5} and above ($N = 897$) is \num{-0.27} and for samples with a confidence score of \num{0.99} and above ($N=712$) is \num{-6.8e-2}.
Note, although we evaluated the AI error relative to the bottom left corner of the image, this result is the same irrespective of the chosen origin because changing the origin amounts to adding a constant vector to $\mathbf{x}_g$ and $\mathbf{x}_b$ that ends up canceling out when calculating $\mathbf{R}_{bg}$.

As with the confidence score, we have examined the relationship between $\Vert\mathbf{R}_{bg}\Vert$ and the IoU associated with the prediction of the object detection algorithm. Figure \ref{rbgvsiour1000m} shows a modest negative linear correlation between $\Vert\mathbf{R}_{bg}\Vert$ and the IoU; the correlation coefficient for the entire dataset is \num{-0.67}. This finding is to be expected since the closer the IoU is to \num{1}, the better the predicted bounding box matches the ground truth.
However, we have also considered $\text{IoU}\ge 0.5$ in our study, and the correlation coefficient is \num{-0.32}. This change can be explained from the scatter plot, which appears more horizontal between the IoU levels of \num{0.5} and \num{0.9}, indicating less correlation. This conjecture is confirmed by the two order of magnitude reduction in absolute value of the covariance to \num{-4.4e-3}; the covariance of the entire sample is \num{-0.10}.
Likewise, when the confidence score is restricted to \num{0.99} and above, the correlation coefficient changes to \num{-0.45}, again, due to a two order of magnitude drop in absolute value of the covariance to \num{-8.4e-3}. This finding is not surprising either since most observations of the AI error with confidence scores at the \num{0.99} level and above (\num{712} observations) have IoUs of \num{0.5} and above. 

In Figure \ref{rbgvsarear1000m}, while it appears that there could be some modest positive correlation between $\Vert\mathbf{R}_{bg}\Vert$ and the detected bounding box area, a correlation coefficient of \num{0.15} does not support this in the linear case. However, when considering only those observations having an IoU of \num{0.5} and above, the correlation coefficient increases to by a factor of \num{3.3} to \num{0.64}, which supports a modest correlation. This is because the standard deviation of $\Vert\mathbf{R}_{bg}\Vert$ reduces a larger amount, by \SI{85}{\percent} from \num{0.85} to \num{0.13}, than the covariance and standard deviation of the detected bounding box area -- the covariance drops only \SI{38}{\percent} from \num{9.6e-3} to \num{5.9e-3} and the standard deviation of the detected box area reduces just \SI{13}{\percent} from \num{7.4e-2} to \num{7.3e-2}. As for those observations with a confidence score of \num{0.99} and above, the correlation coefficient increases by \SI{40}{\percent} to \num{0.21}, which is still not indicative of a strong correlation. Despite this result, these findings support some correlation, but they may warrant further investigation into nonlinear relationships.

The results of this analysis indicate that the IoU may be a useful metric for assessing the impacts on the accuracy in moving a turret system to an aimpoint measured from object detectors developed from AI methods. The predicted bounding box area could be another possible metric, but further investigation is needed into nonlinear correlations with the AI error. While the results do not support a strong linear correlation between the confidence score and the AI error, this does not necessarily rule out the confidence score as a measure of accuracy in moving the turret system to an aimpoint. This can be inferred from Figure \ref{iouvscscorer1000m}, which shows a modest correlation of the IoU with the confidence score. The correlation coefficient for the entire dataset is \num{0.61}, while considering only those samples with IoU of \num{0.5} and above reduces the correlation coefficient by just \SI{1.6}{\percent} to \num{0.60}. This modest positive correlation with the IoU suggests that the confidence score could be an indicator of turret movement accuracy. The results in Section \ref{metricsAndPh} provide further support for the confidence score in this regard. Moreover, as already mentioned, the correlation coefficient is limited to linear relationships and does not rule out any nonlinear relationship. But the other issue regarding the confidence score with this dataset is that the observations of the AI error are heavily weighted towards detections with high confidence scores: \SI{79}{\percent} of the samples have confidence scores of \num{0.8} and above. Given that the AI error varies significantly for these observations, ranging from \SI{3.0e-4}{mils} to \SI{6.1}{mils}, we believe the low correlation coefficient is attributed to not having enough samples at lower confidence scores.

\subsection{Analysis of the Three Error Distributions}
\label{threeErrorDistributions}
\begin{figure}
    \begin{subfigure}{0.24\textwidth}
    \includegraphics[width=\linewidth]{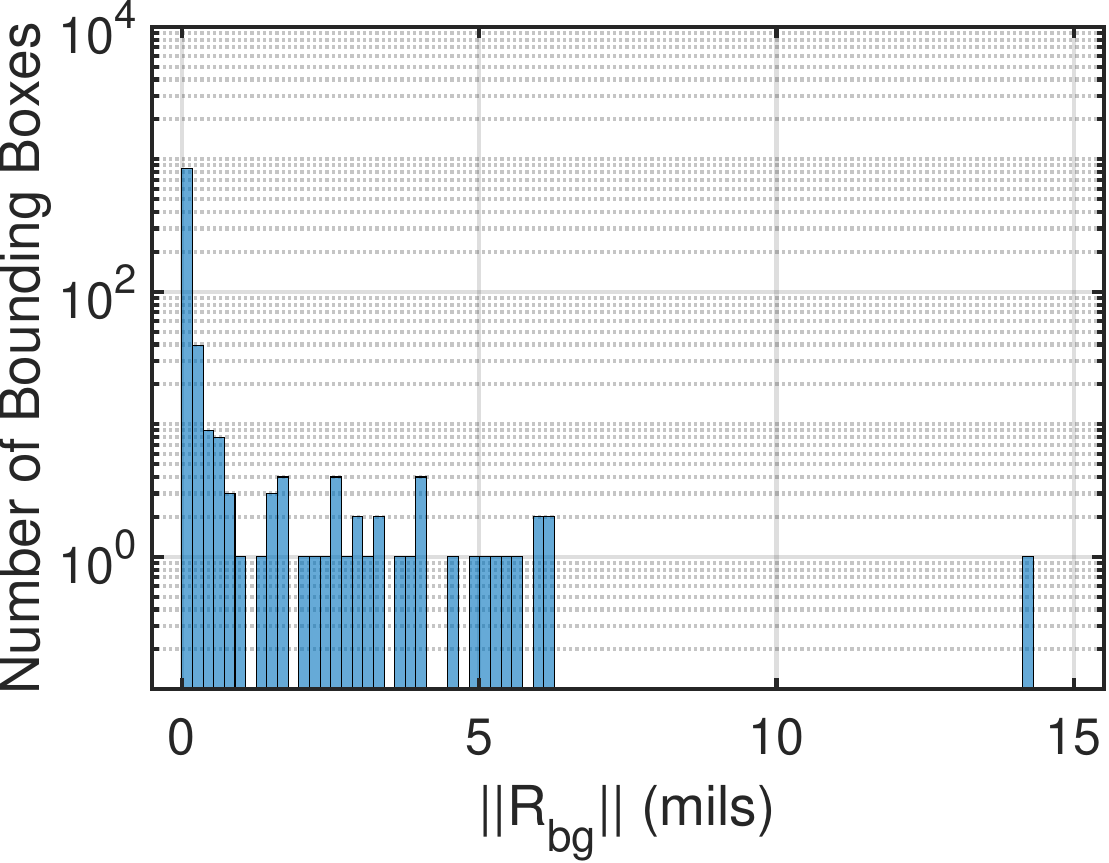}
    \caption{}
    \label{histogramrbgr1000mbtlc}
    \end{subfigure}
    \begin{subfigure}{0.24\textwidth}
        \centering
    \includegraphics[width=\linewidth]{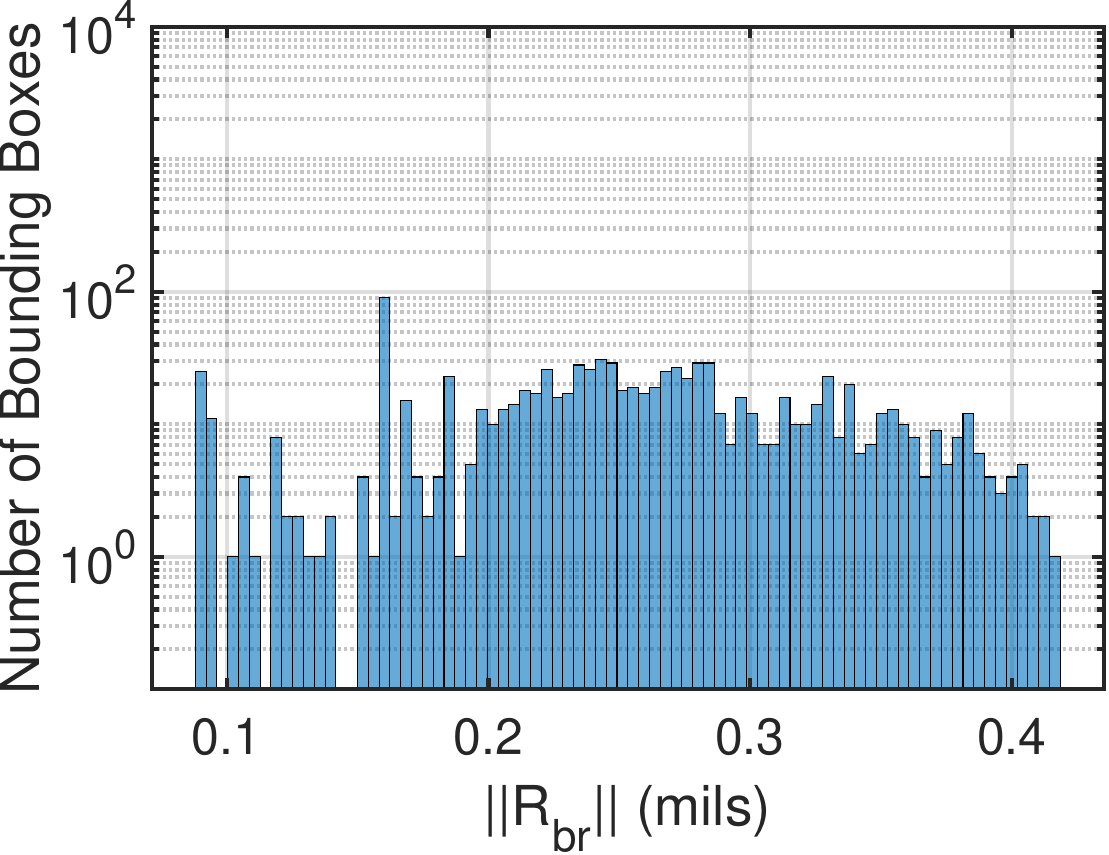}
        \caption{}
        \label{histogramrbrr1000mbtlc}
    \end{subfigure}
    \begin{subfigure}{0.24\textwidth}
        \centering
        \includegraphics[width=\linewidth]{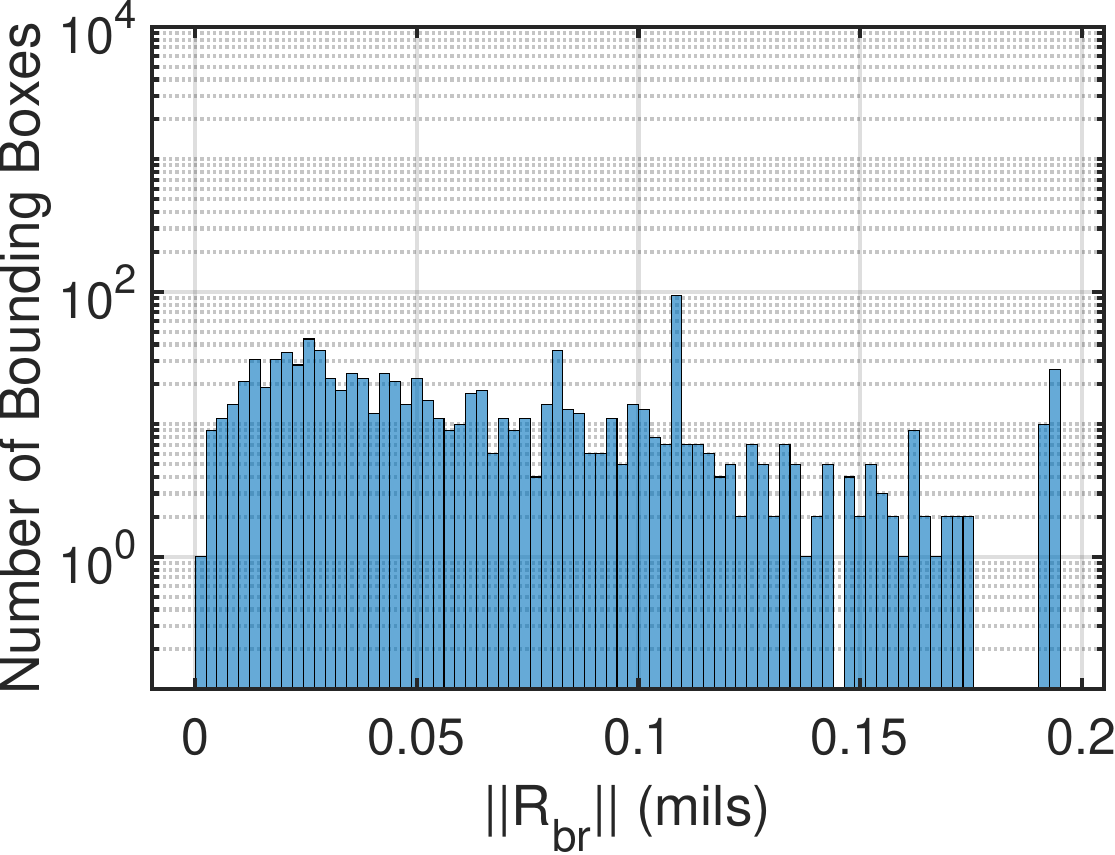}
        \caption{}
        \label{histogramrbrr1000mcenter}
    \end{subfigure}
    \begin{subfigure}{0.24\textwidth}
        \centering
        \includegraphics[width=\linewidth]{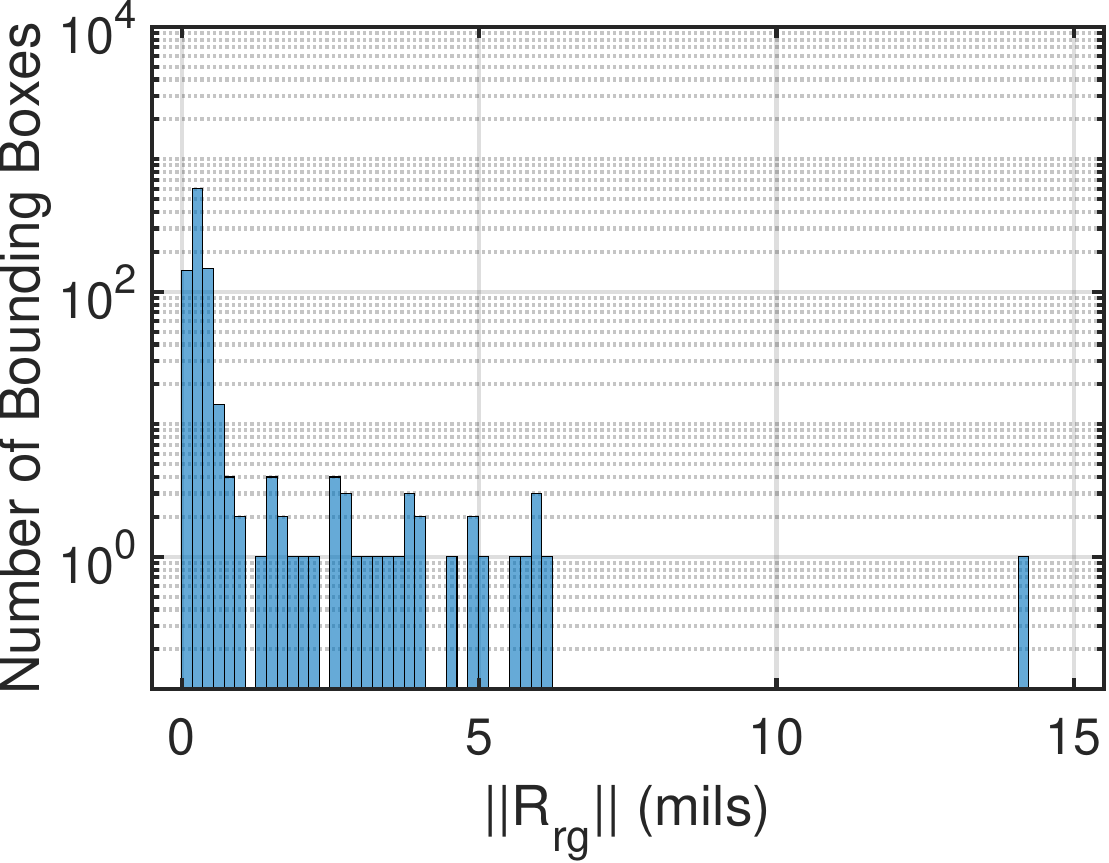}
        \caption{}
        \label{histogramrrgr1000mbtlc}
    \end{subfigure}
    \caption{Histograms of the AI error, controller error and total error. The AI error histogram is shown in (\subref{histogramrbgr1000mbtlc}); while the controller error histogram is shown for targeting starting from the bottom left of the image in (\subref{histogramrbrr1000mbtlc}) and starting from the center of the image in (\subref{histogramrbrr1000mcenter}); and the total error histogram is shown for targeting starting from the bottom left of the image in (\subref{histogramrrgr1000mbtlc}). We omit the histogram of the total error sample with targeting starting from the center of the image for clarity as it looks similar to the histogram in (\subref{histogramrrgr1000mbtlc}).
    The total number of observations in each plot is $N=955$.}
    \label{histogramrbrr1000m}
\end{figure}

As another part of our analysis of AI impacts on the movement error of our controlled gun turret system, we examine the three error sources on histograms beginning with the AI error $\Vert\mathbf{R}_{bg}\Vert$ in Figure \ref{histogramrbgr1000mbtlc}. As with the plots in Section \ref{metricsAndAiError}, the plot is the same irrespective of the choice of reference origin used to evaluate the error. The histogram shows the level of error ranges between \SI{0}{mils} and \SI{14}{mils} assuming each target is at a range of \SI{1000}{\meter}; the sample mean is \SI{0.22}{mils} and the standard deviation is \SI{0.85}{mils}. The importance of this plot is that it provides evidence of the level of error contributed by the object detection model to the controlled gun turret system assuming a fixed range for each target; this quantifies the impact on weapon system accuracy and is useful for the development of error budgets accounting for AI in weapon systems.

In Figure \ref{histogramrbrr1000mbtlc} and Figure \ref{histogramrbrr1000mcenter}, the histograms of the controller error $\Vert\mathbf{R}_{br}\Vert$ are notably different than histogram of the AI error. The differences are observed mainly in the size of the error and the sample statistics. The scale ranges from \SI{0}{mils} to \SI{0.42}{mils} in Figure \ref{histogramrbrr1000mbtlc} for aiming starting from the bottom left corner of the image, and from \SI{0}{mils} to \SI{0.20} {mils} in Figure \ref{histogramrbrr1000mcenter} for aiming starting from the center for the image; this is roughly a \SI{54}{\percent} decrease in the range of error. Regarding the statistics of each histogram, the mean decreases by \SI{74}{\percent} from \SI{0.25}{mils} to \SI{6.7e-2}{mils} while the standard deviation decreases by \SI{35}{\percent} from \SI{7.4e-2}{mils} to \SI{4.8e-2}{mils} when turret movement starts from the center of the image. These decreases are to be expected since, on average, the gun barrel moves a shorter distance to reach the aimpoint when starting from the center of the image and controller errors are proportional to the aimpoint. Additionally, we also observe that the standard deviation of the controller error in both targeting scenarios is \num{12} to \num{18} times smaller than the standard deviation of the AI error sample. This finding suggests that the AI error is an important error to account for in the accuracy in moving the controlled turret to the aimpoint. 

The histogram of the total error $\Vert\mathbf{R}_{rg}\Vert$ sample supports this claim, which is evident in Figure \ref{histogramrrgr1000mbtlc}, since it appears to closely resemble the histogram of the AI error sample in Figure \ref{histogramrbgr1000mbtlc}. As with the AI error sample, the level of error ranges between \SI{0}{mils} and \SI{14}{mils} for both targeting scenarios. Nevertheless, there are significant changes in the sample statistics, mainly with respect to the sample mean. For instance, the mean of the total error sample is \SI{0.41}{mils} when aiming starts from the bottom left corner of the image, which is \SI{86}{\percent} larger than the corresponding mean of the AI error sample. But this drops to \SI{0.26}{mils} when aiming starts from the center of the image, which is \SI{17}{\percent} larger than the mean of the AI error sample and \SI{37}{\percent} smaller than the mean of the total error sample when starting to aim from the bottom left corner of the image. These changes in the mean are due to the fact that the shot is not exactly on the aim point at the firing time and that the total error includes the controller error. This result is to be expected since, generally, the error between the aimpoint and shot, quantified by $\Vert\mathbf{R}_{br}\Vert$, will be close to \num{0} if the firing time is near the settling time of the controller. But overall, these differences in the mean of both total error samples simply shift the observations to the right or left on the histograms; they do not affect how the observations cluster around the mean. The determining statistic in that regard, and ultimately in the resemblance, is the standard deviation of both total error samples, which did not change much compared to the corresponding standard deviation of the AI error sample. We observe a decrease of only \SI{3.5}{\percent} to \SI{0.82}{mils} when aiming starts from the bottom left corner of the image and no change when aiming starts from the center of the image. This finding is reasonable and can be explained from the variance of the total error $\sigma_{rg}^2$ in the following way.

By the triangle equality, the total error satisfies the bound: $\Vert\mathbf{R}_{rg}\Vert\le\Vert\mathbf{R}_{bg}\Vert+\Vert\mathbf{R}_{br}\Vert$. If we define the variable $\varepsilon = \Vert\mathbf{R}_{bg}\Vert+\Vert\mathbf{R}_{br}\Vert - \Vert\mathbf{R}_{rg}\Vert$, the total error is $
    \Vert\mathbf{R}_{rg}\Vert = \Vert\mathbf{R}_{bg}\Vert + \Vert\mathbf{R}_{br}\Vert - \varepsilon
    $.
Using this identity, the variance of the total error can then be expanded as:
\begin{equation}
    \label{varianceRrg}
    \sigma_{rg}^2 = \sigma_{bg}^2 + \sigma_{br}^2 + \sigma_{\varepsilon}^2 + 2\text{Cov}(\Vert\mathbf{R}_{bg}\Vert,\Vert\mathbf{R}_{br}\Vert) - 2\text{Cov}(\Vert\mathbf{R}_{bg}\Vert,\varepsilon) - 2\text{Cov}(\Vert\mathbf{R}_{br}\Vert,\varepsilon), 
\end{equation}
where $\sigma_{bg}^2$, $\sigma_{br}^2$ and $\sigma_{\varepsilon}^2$ are the variances of the AI error, controller error and $\varepsilon$, respectively; see the Appendix for a derivation of this result. 

A reasonable approximation to $\sigma_{rg}^2$ can be made by the sum $\sigma_{bg}^2+\sigma_{br}^2$ in \eqref{varianceRrg} if the sum of the remaining terms is smaller by at least an order of magnitude or more in absolute value. This approximation is valid for our error data since the sum $\sigma_{bg}^2+\sigma_{br}^2$ is $\num{0.73}$ in both targeting scenarios, which is a factor of \num{12} larger than the absolute value of the remaining terms (\num{6.0e-2}) in \eqref{varianceRrg} when aiming starts from the bottom left corner of the image and a factor of \num{83} larger than the absolute value of this sum (\num{8.9e-3}) when aiming starts from the center of the image.  Consequently, the sum $\sigma_{bg}^2 + \sigma_{br}^2$ of \num{0.73} is only \SI{9}{\percent} larger than the true value of \num{0.67} for $\sigma_{rg}^2$ when turret movement starts from the bottom left corner of the image and just \SI{1.4}{\percent} larger than the true value of \num{0.72} for $\sigma_{rg}^2$ when turret movement starts from the center of the image.  

Based on this approximation, we see that the variance of the total error will be close to the variance of the AI error if the variance of the controller error is sufficiently small. From our error data this is indeed the case: the variance of AI error sample is \num{0.72}, while the variance of the controller error sample is \num{5.4e-3} for aiming starting from the bottom left corner of the image and \num{2.3e-3} for aiming starting from the center of the image. In both cases, the variance of the controller error sample is at least two orders of magnitude smaller than the variance of the AI error sample. Thus, most of the total error comes from the AI error, which explains why the standard deviations of the total error do not change much in either targeting scenario, and also why the histograms of the total error and AI error closely resemble each other.

The results of this analysis indicate that the AI error $\Vert\mathbf{R}_{bg}\Vert$ can significantly impact the accuracy in moving the controlled gun turret system to the aimpoint. The statistics of the AI error sample provide a preliminary quantification of the error added to a targeting system by an AI technology assuming a fixed range for each target. The other important result is that the distribution of the total error $\Vert\mathbf{R}_{rg}\Vert$ will closely resemble the distribution of the AI error if the variance of the controller error $\Vert\mathbf{R}_{br}\Vert$ is sufficiently smaller than the variance of the AI error. In this case, the AI error approximates the total error very well. Another finding is that starting from the center of the image produces smaller averages in the controller error and the total error, but only a smaller variance in the controller error. This is because $\Vert\mathbf{R}_{bg}\Vert$ is strictly the error in the AI prediction while the total error  contains the error from the AI prediction and the controller. Note that, in general, this error is not additive by the triangle inequality: $\Vert\mathbf{R}_{rg}\Vert\le\Vert\mathbf{R}_{bg}\Vert+\Vert\mathbf{R}_{br}\Vert$. However, the result of lower variance for the controller error is under the assumption of no uncertainty in the measurement of the aimpoint and outputs of the feedback system. When considering these sources of uncertainty, this result would not necessarily hold since the controller cannot completely attenuate the effects of the uncertainty.

\subsection{Relationship between Object Detection Metrics and the Probability of a Hit}
\label{metricsAndPh}
In this section, we describe the results from examining correlations between object detection performance metrics and the probability of hit. For this analysis, we have sampled five smaller datasets from the object detector output dataset analyzed in Section \ref{metricsAndAiError} according to five IoU ranges: $[0.0,0.6]$, $[0.6,0.7]$, $[0.7,0.8]$, $[0.8,0.9]$ and $[0.9,1]$; the datasets are labeled A, B, C, D and E, accordingly. When performing the sampling, we have imposed the condition that the area of the ground truth bounding box area remain within \SI{2.5}{\percent} of the largest area of all the ground truth bounding boxes. The value of \SI{2.5}{\percent} has been chosen to maintain a sufficient sample size. Additionally, we exclude detections with IoUs of \num{0} as these account for only \SI{0.30}{\percent} of the detections in the object detector output dataset and occur with low confidence scores (less than \num{0.4}). This sampling procedure has led to a sample size of \num{747} detections among these five datasets which is \SI{78}{\percent} of the detections in the object detector output dataset.

The rationale for this sampling procedure is to keep the target size similar among the images. This is because the size of the ground truth bounding box should not vary much if the range is the same for each target. However, the range of the target in the images examined in this study is not fixed. For instance, there are images where the UAV is within roughly \SI{1}{\meter} of the camera and appears quite large; the ground truth bounding boxes have areas ranging from \SI{10}{\percent} to \SI{20}{\percent} of the area of the image in these cases. In one of the images with a large target, there are \num{11} detected bounding boxes with IoUs less than \num{0.5} that have areas which are less than \SI{25}{\percent} of the ground truth area and also lie within the ground truth bounding box; these observations comprise \SI{19}{\percent} of the \num{58} detected bounding boxes with IoUs below \num{0.5}. Consequently, a probability of a hit of \SI{100}{\percent} is observed for these detections despite the low IoU. The problem with this is that the dataset is heavily weighted towards smaller targets: \SI{89}{\percent} of the ground truth bounding boxes have areas less than \SI{1}{\percent} of the area of the image. Thus, not accounting for these larger ground truth bounding boxes can lead to conflicting results. This is because \SI{88}{\percent} of the detections with IoUs above \num{0.5} occur in images with smaller ground truth bounding boxes, i.e., having areas at most \SI{1}{\percent} of the area of the image, and end up achieving a lower probability of a hit.

With these five datasets, we evaluate the following metrics according to the steps outlined in Section \ref{metricsForObjectDetectors}: AP50, AP75, AP@50:5:95, AR1 and AR10; we have adopted the all-point interpolation method to evaluate the three AP metrics. Afterwards, we perform numerical targeting simulations with our controlled gun turret model against each detected target in the datasets. The probability of the hit is evaluated at the end of every simulation by applying equation \eqref{phEqn} to the ground-truth bounding box data, which is used to determine the limits of integration, and the results are grouped by each dataset and averaged by the number of detections per dataset. This procedure is done for six uniformly spaced ranges from \SI{500}{\meter} to \SI{3000}{\meter} with turret movement starting from the bottom left corner of the image and repeated for turret movement starting from the center of the image. 

When performing the simulations with each dataset, we fix a range and add estimates of the errors arising in the projectile trajectory calculated in TAM \citep{bunn1993tank} to form the means and variances of the bivariate impact distribution, as described in Section \ref{evaluatePh}, and evaluate the probability of a hit. These error values are the total biases and random errors in the azimuth and elevation directions as a function of target range output by TAM using the default inputs for each error source. The exception is that we set the lay error input to \num{0} since we use the standard deviation of the controller error, described in the Appendix, as an estimate of the random component of the error in moving the gun turret to the aimpoint.
\begin{table}
    \centering
      \caption{Performance metrics for the five datasets sampled from the output dataset of the object detection model evaluated on UAV imagery. The second column reports the number of detections, while columns three and four report the average confidence score and average IoU and average normalized ground truth bounding box area in each dataset.}
    \begin{tabular}{ccccccccc}
    \hline
      \textbf{Dataset} & $N_{D}$ & \textbf{Confidence Score} & \textbf{IoU} & \textbf{AP50} & \textbf{AP75} & \textbf{AP@50:5:95} & \textbf{AR1} & \textbf{AR10} \\
         \hline
         A & \num{50} & \num{0.32} & \num{0.50} & \num{0.40} & \num{0.00} & \num{0.051} & \num{0.067} & \num{0.070} \\
         B & \num{95} & \num{0.47} & \num{0.66} & \num{0.99} & \num{0.00} & \num{0.34} & \num{0.32} & \num{0.32}  \\
         C & \num{127} & \num{0.76} & \num{0.75} & \num{1.0} & \num{0.33} & \num{0.53} & \num{0.50} & \num{0.50} \\
         D & \num{248} & \num{0.99} & \num{0.85} & \num{1.0} & \num{1.0} & \num{0.73} & \num{0.71} & \num{0.71} \\
         E & \num{224} & \num{1.0} & \num{0.94} & \num{1.0} & \num{1.0} & \num{0.91} & \num{0.88} & \num{0.88} \\
         \hline
    \end{tabular}
    \label{aparresults}
\end{table}

\begin{figure}
    \begin{subfigure}{0.5\textwidth}
        \centering
        \includegraphics[width=0.9\linewidth]{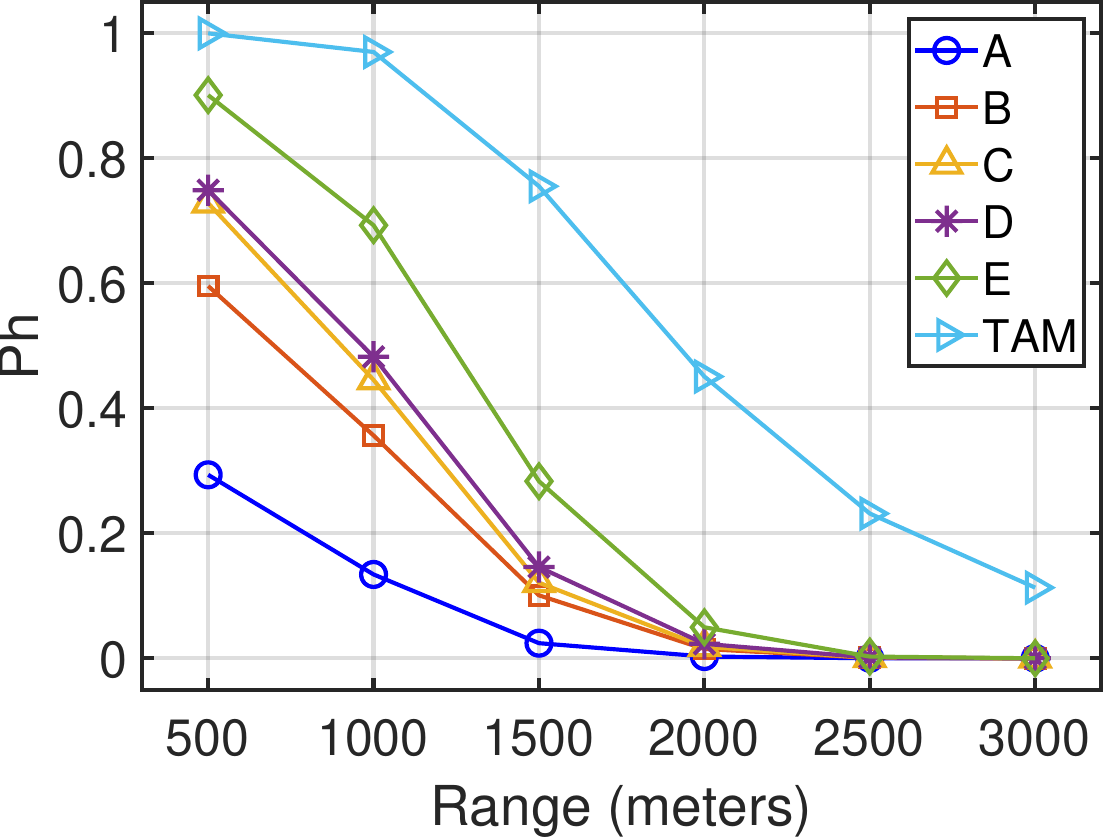}
        \caption{}
        \label{metricsRangePhPlotBtlc}
    \end{subfigure}%
    \begin{subfigure}{0.5\textwidth}
        \centering
        \includegraphics[width=0.9\linewidth]{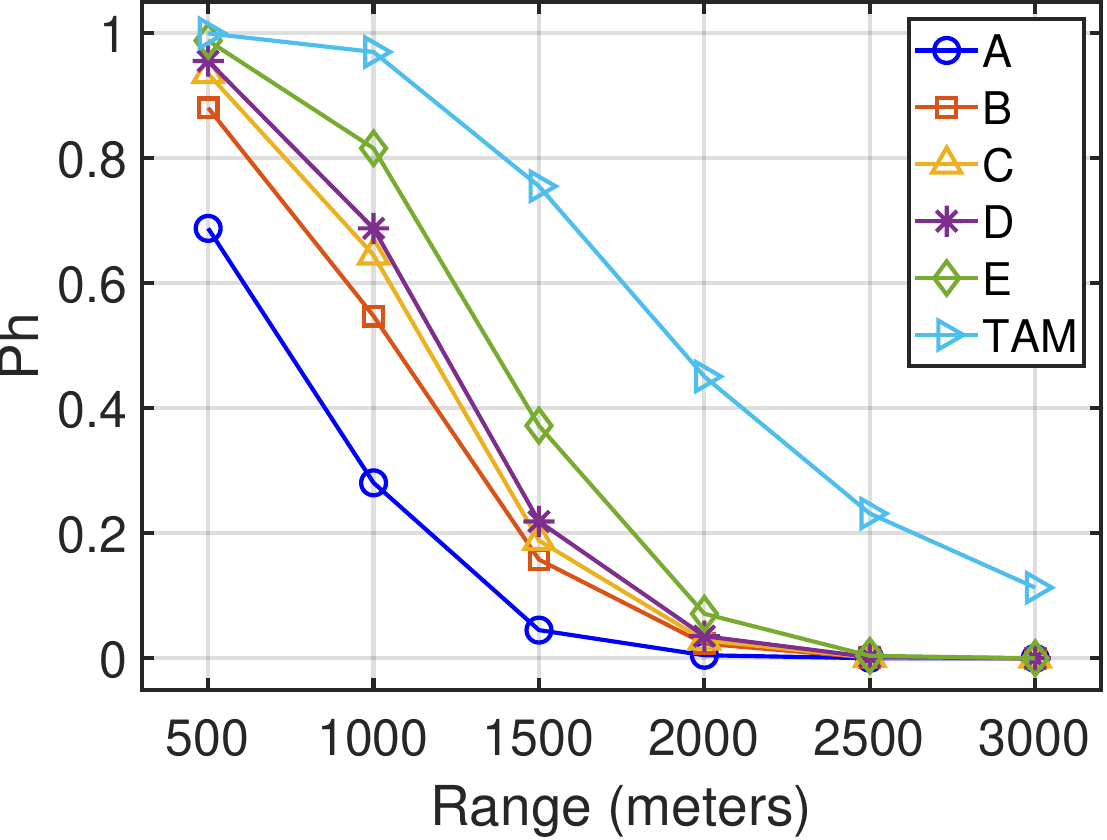}
        \caption{}
        \label{metricsRangePhPlotCenter}
    \end{subfigure}
    \caption{Plots of the probability of a hit as a function of target range for the five datasets sampled from the output dataset of the object detection model evaluated on UAV imagery; each curve corresponds to a dataset. We plot the probability of a hit output from TAM for comparison. In (a), targeting starts from the bottom left corner of the image and in (b), targeting starts from the center of the image.}
    \label{metricsRangePhPlot}
\end{figure}

In comparison of the results in Table \ref{aparresults} with both plots in Figure \ref{metricsRangePhPlot}, we observe that all metrics are positively correlated with the probability of a hit for all ranges considered. However, the correlation is not strong beyond a range of \SI{2000}{\meter} since the probability of a hit drops below \SI{10}{\percent}; at this point, the projectile motion errors calculated from TAM become the dominate sources of error. Additionally, some metrics are less sensitive to changes in the probability of a hit than others, even at ranges below \SI{2000}{\meter}, namely AP50 and AP75.  While these metrics do show a positive correlation with the probability of a hit within these ranges, interpretation of this correlation is limited due to small changes in these metrics between the datasets. This outcome is not surprising since the detections in four of the datasets (B, C, D and E) have an average IoU above \num{0.6} while the detections in three of the datasets (C, D and E) have an average IoU of \num{0.75} and above. The results in Table \ref{aparresults} support this finding: AP50 increases by a factor of \num{1.5} between datasets A and B, but then increases by only \SI{1}{\percent} between datasets B and C and remains at the highest scoring of \num{1.0} thereafter. On the other hand, AP75 shows some more variation increasing by \num{0.33} from \num{0.0} between dataset B and C and then by a factor of \num{2} between dataset C and D to also reach the highest scoring of \num{1.0}. Consequently, the five datasets are highly insensitive to AP50 and moderately insensitive to AP75.

Another important outcome seen in Figure \ref{metricsRangePhPlotBtlc} and Figure \ref{metricsRangePhPlotCenter} is that the probability of a hit decreases with increasing range for each dataset; this is to be expected and is consistent with the probability of a hit output from TAM. Moreover, the probability of a hit at each range considered is smaller than the probability of a hit calculated by TAM as seen in both Figure \ref{metricsRangePhPlotBtlc} and Figure \ref{metricsRangePhPlotCenter}. The decrease occurs because the lay error input is set to \num{0} and the AI error is an additional error source that is not accounted for in TAM. Additionally, the probability of a hit is higher in Figure \ref{metricsRangePhPlotCenter} for the five datasets at all ranges considered than in Figure \ref{metricsRangePhPlotBtlc}. This result is not surprising either because, as noted in Section \ref{threeErrorDistributions}, the gun barrel moves less distance on average when starting from the center of the image and controller errors are proportional to the aimpoint. 
\section{Conclusions}
In this paper, we assess the impacts of an AI technology on the accuracy of a controlled gun turret system given target location as input. For this purpose, we use output data from an AI object detection model evaluated on UAV imagery to determine the azimuth and elevation aimpoints for the turret system in numerical targeting simulations. As part of our assessment, we analyze the bounding box centroid error, referred to as the AI error, and the resulting errors in moving the controlled gun turret system to the aimpoint. We then perform additional numerical simulations of our controlled gun turret system using target location data from five datasets sampled from the given output dataset of the object detector. We evaluate the probability of a hit at the end of each simulation and study the correlations with several performance metrics for object detectors.

The analysis of the AI error and aiming error data of the controlled gun turret system given target location from AI data as input reveals some important outcomes. The first is that the level of AI error observed in our study provides evidence that AI technologies, such as object detectors, are significant sources of error to a targeting system, and serves as a preliminary quantification of the impacts on the accuracy of the system needed for the development of error budgets accounting for AI. Furthermore, the results suggest that metrics such as the IoU and detected bounding box area could be predictors of the level of error introduced to the targeting system by an AI technology; this is indicated by a modest negative correlation of the IoU with the AI error and a modest positive correlation of the bounding box area with the AI error when considering samples having IoUs of \num{0.5} and above. Another important outcome is that the AI error is the most contributing source of error in moving the gun turret to the aimpoint provided the uncertainty in the controller error is at least two orders of magnitude smaller than the uncertainty in the AI error; this is supported by the observed variances of the AI and controller errors in this study. The third outcome of our analysis is that starting to move the gun barrel from the center of the image rather than the bottom left corner leads to smaller averages in the total error and controller error, and also to a smaller variance in the controller error; but this may not always be the case in a physical system where outputs are measured from sensors and inputs have added uncertainty arising from image quality defects that is passed to the feedback system after image processing by the object detection model.

From the results of our numerical experiment evaluating object detection metrics and the probability of a hit, we observe that the confidence score, IoU, AP and AR are positively correlated with the probability of a hit for all ranges considered in this study. However, beyond ranges of \SI{2000}{\meter}, the correlation is not very strong as the probability of a hit drops well below \SI{10}{\percent}. This is because the errors accounted for by TAM dominate both the AI and controller errors at these ranges. Moreover, the interpretation of this correlation with the metrics AP50 and AP75 is limited, even at ranges below \SI{2000}{\meter}, owing to the datasets being insensitive to the IoU thresholds of \num{0.5} and \num{0.75}. This is because all five datasets have average IoUs of \num{0.5} and above while three of the datasets have average IoUs of \num{0.75} and above. We also observe higher probabilities of a hit when turret movement starts from the center of the image, which is expected since the gun barrel moves a shorter distance on average. These findings suggest that the confidence score, IoU, AP@50:5:95 and AR are predictors of high performance of object detectors in ATR algorithms.

\section*{Acknowledgment}
The authors would like to gratefully acknowledge the DEVCOM Analysis Center (DAC) for the support and guidance on this work. Additionally, we would like to acknowledge Eamon Conway at Northeastern University for sharing with us the output data of the object detection model that is studied in this paper. Lastly, we acknowledge Rifat Sipahi, Paola Kefallinos, Tony Smoragiewicz and Matthew Tortolani at Northeastern University for prodiving us the updated MATLAB version of TAM used to generate the additional errors arising from projectile motion in our simulations.

This research was sponsored by the DEVCOM Analysis Center and was accomplished under Contract Number W911QX--23--D0002. The views and conclusions contained in this document are those of the authors and should not be interpreted as representing the official policies, either expressed or implied, of the Army Research Office or the U.S. Government. The U.S. Government is authorized to reproduce and distribute reprints for Government purposes notwithstanding any copyright notation herein.
\bibliography{References}

\begin{thebibliography}{50}
\providecommand{\natexlab}[1]{#1}
\providecommand{\url}[1]{\texttt{#1}}
\expandafter\ifx\csname urlstyle\endcsname\relax
  \providecommand{\doi}[1]{doi: #1}\else
  \providecommand{\doi}{doi: \begingroup \urlstyle{rm}\Url}\fi

\bibitem[Bendat and Piersol(2010)]{Bendat2010RandomData}
J.~S. Bendat and A.~G. Piersol.
\newblock \emph{Random Data: Analysis and Measurement Procedures}, chapter~6,
  pages 173--199.
\newblock Wiley, Hoboken, NJ, Jan. 2010.
\newblock ISBN 9780470248775.
\newblock \doi{10.1002/9781118032428}.

\bibitem[Bhanu(1986)]{Bhanu1986Automatic}
B.~Bhanu.
\newblock Automatic target recognition: State of the art survey.
\newblock \emph{IEEE Transactions on Aerospace and Electronic Systems},
  AES-22\penalty0 (4):\penalty0 364--379, July 1986.
\newblock ISSN 0018-9251.
\newblock \doi{10.1109/taes.1986.310772}.

\bibitem[Blasch et~al.(2020)Blasch, Majumder, Zelnio, and
  Velten]{Blasch2020Review}
E.~Blasch, U.~K. Majumder, E.~G. Zelnio, and V.~J. Velten.
\newblock Review of recent advances in ai/ml using the mstar data.
\newblock In E.~Zelnio and F.~D. Garber, editors, \emph{Algorithms for
  Synthetic Aperture Radar Imagery XXVII}. SPIE, May 2020.
\newblock \doi{10.1117/12.2559035}.

\bibitem[Bunn(1993)]{bunn1993tank}
F.~L. Bunn.
\newblock The tank accuracy model.
\newblock Technical Report ARL-MR-48, Army Research Lab, Aberdeen Proving
  Ground, MD, Mar. 1993.

\bibitem[Carlstedt(2021)]{carlstedt2021modelling}
A.~Carlstedt.
\newblock Modelling of electromechanical motors for turret and barrel control
  in main battle tanks.
\newblock Master's thesis, KTH Royal Institute of Technology, Stockholm,
  Sweden, 2021.

\bibitem[Chen et~al.(2016)Chen, Wang, Xu, and Jin]{Chen2016Target}
S.~Chen, H.~Wang, F.~Xu, and Y.~Jin.
\newblock Target classification using the deep convolutional networks for sar
  images.
\newblock \emph{IEEE Transactions on Geoscience and Remote Sensing},
  54\penalty0 (8):\penalty0 4806--4817, Aug. 2016.
\newblock ISSN 1558-0644.
\newblock \doi{10.1109/tgrs.2016.2551720}.

\bibitem[Christiansen et~al.(2014)Christiansen, Steen, Jørgensen, and
  Karstoft]{Christiansen2014Automated}
P.~Christiansen, K.~Steen, R.~Jørgensen, and H.~Karstoft.
\newblock Automated detection and recognition of wildlife using thermal
  cameras.
\newblock \emph{Sensors}, 14\penalty0 (8):\penalty0 13778--13793, July 2014.
\newblock ISSN 1424-8220.
\newblock \doi{10.3390/s140813778}.

\bibitem[El-Darymli et~al.(2016)El-Darymli, Gill, Mcguire, Power, and
  Moloney]{ElDarymli2016Automatic}
K.~El-Darymli, E.~W. Gill, P.~Mcguire, D.~Power, and C.~Moloney.
\newblock Automatic target recognition in synthetic aperture radar imagery: A
  state-of-the-art review.
\newblock \emph{IEEE Access}, 4:\penalty0 6014--6058, 2016.
\newblock ISSN 2169-3536.
\newblock \doi{10.1109/access.2016.2611492}.

\bibitem[Everingham et~al.(2014)Everingham, Eslami, Gool, Williams, Winn, and
  Zisserman]{Everingham2014Pascal}
M.~Everingham, S.~M.~A. Eslami, L.~V. Gool, C.~K.~I. Williams, J.~Winn, and
  A.~Zisserman.
\newblock The pascal visual object classes challenge: A retrospective.
\newblock \emph{International Journal of Computer Vision}, 111\penalty0
  (1):\penalty0 98--136, June 2014.
\newblock \doi{10.1007/s11263-014-0733-5}.

\bibitem[Fan and Liu(2019)]{Fan2019Challenges}
J.~Fan and J.~Liu.
\newblock The challenges and some thinking for the intelligentization of
  precision guidance atr.
\newblock In \emph{Artificial Intelligence and Machine Learning in Defense
  Applications}. SPIE, Sept. 2019.
\newblock \doi{10.1117/12.2532261}.

\bibitem[Fernández-Caballero et~al.(2014)Fernández-Caballero, López, and
  Serrano-Cuerda]{FernandezCaballero2014Thermal}
A.~Fernández-Caballero, M.~López, and J.~Serrano-Cuerda.
\newblock Thermal-infrared pedestrian roi extraction through thermal and motion
  information fusion.
\newblock \emph{Sensors}, 14\penalty0 (4):\penalty0 6666--6676, Apr. 2014.
\newblock ISSN 1424-8220.
\newblock \doi{10.3390/s140406666}.

\bibitem[Franklin et~al.(2015)Franklin, Powell, and
  Emami-Naeini]{FranklinCh6FreqResp}
G.~F. Franklin, J.~D. Powell, and A.~Emami-Naeini.
\newblock \emph{Feedback Control of Dynamic Systems}, chapter~6, pages
  308--432.
\newblock Pearson, Upper Saddle River, NJ, 7 edition, 2015.
\newblock ISBN 9780134685717.

\bibitem[Gardony et~al.(2022)Gardony, Okano, Hughes, Kim, Renshaw, and
  Sipolins]{Gardony2022Aided}
A.~L. Gardony, K.~Okano, G.~I. Hughes, A.~J. Kim, K.~T. Renshaw, and
  A.~Sipolins.
\newblock Aided target recognition visual design impacts on cognition in
  simulated augmented reality.
\newblock \emph{Frontiers in Virtual Reality}, 3, Sept. 2022.
\newblock ISSN 2673-4192.
\newblock \doi{10.3389/frvir.2022.982010}.

\bibitem[Gilmore(1984)]{Gilmore1984Artificial}
J.~F. Gilmore.
\newblock Artificial intelligence in automatic target recognizers: Technology
  and timelines.
\newblock In \emph{Applications of Digital Image Processing VII}, volume 504,
  pages 34--39. SPIE, Dec. 1984.
\newblock \doi{10.1117/12.944843}.

\bibitem[Girshick(2015)]{Girshick2015Fast}
R.~Girshick.
\newblock Fast {R-CNN}.
\newblock In \emph{2015 {IEEE} International Conference on Computer Vision
  ({ICCV})}. {IEEE}, Dec. 2015.
\newblock \doi{10.1109/iccv.2015.169}.

\bibitem[Girshick et~al.(2014)Girshick, Donahue, Darrell, and
  Malik]{Girshick2014Rich}
R.~Girshick, J.~Donahue, T.~Darrell, and J.~Malik.
\newblock Rich feature hierarchies for accurate object detection and semantic
  segmentation.
\newblock In \emph{2014 {IEEE} Conference on Computer Vision and Pattern
  Recognition}. {IEEE}, June 2014.
\newblock \doi{10.1109/cvpr.2014.81}.

\bibitem[He et~al.(2015)He, Zhang, Ren, and Sun]{He2015Spatial}
K.~He, X.~Zhang, S.~Ren, and J.~Sun.
\newblock Spatial pyramid pooling in deep convolutional networks for visual
  recognition.
\newblock \emph{{IEEE} Transactions on Pattern Analysis and Machine
  Intelligence}, 37\penalty0 (9):\penalty0 1904--1916, Sept. 2015.
\newblock \doi{10.1109/tpami.2015.2389824}.

\bibitem[Hollands et~al.(2018)Hollands, Terhaar, and
  Pavlovic]{Hollands2018Effects}
J.~G. Hollands, P.~Terhaar, and N.~J. Pavlovic.
\newblock Effects of resolution, range, and image contrast on target
  acquisition performance.
\newblock \emph{Human Factors: The Journal of the Human Factors and Ergonomics
  Society}, 60\penalty0 (3):\penalty0 363--383, Mar. 2018.
\newblock ISSN 1547-8181.
\newblock \doi{10.1177/0018720818760331}.

\bibitem[Hosang et~al.(2016)Hosang, Benenson, Dollar, and
  Schiele]{Hosang2016What}
J.~Hosang, R.~Benenson, P.~Dollar, and B.~Schiele.
\newblock What makes for effective detection proposals?
\newblock \emph{IEEE Transactions on Pattern Analysis and Machine
  Intelligence}, 38\penalty0 (4):\penalty0 814--830, Apr. 2016.
\newblock ISSN 2160-9292.
\newblock \doi{10.1109/tpami.2015.2465908}.

\bibitem[Idris et~al.(2015)Idris, Hudha, Kadir, and Amer]{idris-hudha-ASCC2015}
A.~M. Idris, K.~Hudha, Z.~A. Kadir, and N.~H. Amer.
\newblock Development of target tracking control of gun-turret system.
\newblock In \emph{2015 10th Asian Control Conference ({ASCC})}. {IEEE}, May
  2015.
\newblock \doi{10.1109/ascc.2015.7244528}.

\bibitem[Jaccard(1901)]{Jaccard1901Etude}
P.~Jaccard.
\newblock Etude comparative de la distribution florale dans une portion des
  alpes et des jura.
\newblock \emph{Bulletin de la Societe Vaudoise des Sciences Naturelles},
  37:\penalty0 547--579, 1901.

\bibitem[Kazemi et~al.(2018)Kazemi, Iranmanesh, and
  Nasrabadi]{Kazemi2018Automatic}
H.~Kazemi, M.~Iranmanesh, and N.~M. Nasrabadi.
\newblock Automatic target recognition using deep convolutional neural
  networks.
\newblock In \emph{Automatic Target Recognition XXVIII}. SPIE, Apr. 2018.
\newblock \doi{10.1117/12.2304643}.

\bibitem[Kechagias-Stamatis and Aouf(2021)]{KechagiasStamatis2021Automatic}
O.~Kechagias-Stamatis and N.~Aouf.
\newblock Automatic target recognition on synthetic aperture radar imagery: A
  survey.
\newblock \emph{IEEE Aerospace and Electronic Systems Magazine}, 36\penalty0
  (3):\penalty0 56--81, Mar. 2021.
\newblock ISSN 1557-959X.
\newblock \doi{10.1109/maes.2021.3049857}.

\bibitem[Liu et~al.(2016)Liu, Anguelov, Erhan, Szegedy, Reed, Fu, and
  Berg]{Liu2016SSD}
W.~Liu, D.~Anguelov, D.~Erhan, C.~Szegedy, S.~Reed, C.-Y. Fu, and A.~C. Berg.
\newblock {SSD}: Single shot {MultiBox} detector.
\newblock In \emph{Computer Vision {\textendash} {ECCV} 2016}, pages 21--37.
  Springer International Publishing, 2016.
\newblock \doi{10.1007/978-3-319-46448-0_2}.

\bibitem[Lyth(2021)]{lyth2021modelling}
M.~Lyth.
\newblock Modeling and evaluation of turret control systems for main battle
  tanks.
\newblock Master's thesis, KTH Royal Institute of Technology, Stockholm,
  Sweden, 2021.

\bibitem[Ma et~al.(2022)Ma, Yang, Sun, Wang, and Wang]{Ma2022adaptive}
Y.~Ma, G.~Yang, Q.~Sun, X.~Wang, and Z.~Wang.
\newblock Adaptive robust feedback control of moving target tracking for all -
  electrical tank with uncertainty.
\newblock \emph{Defence Technology}, 18\penalty0 (4):\penalty0 626--642, Apr.
  2022.
\newblock \doi{10.1016/j.dt.2021.03.005}.

\bibitem[Malmgren-Hansen and
  Nobel-Jorgensen(2015)]{MalmgrenHansen2015Convolutional}
D.~Malmgren-Hansen and M.~Nobel-Jorgensen.
\newblock Convolutional neural networks for sar image segmentation.
\newblock In \emph{2015 IEEE International Symposium on Signal Processing and
  Information Technology (ISSPIT)}, pages 231--236. IEEE, Dec. 2015.
\newblock \doi{10.1109/isspit.2015.7394333}.

\bibitem[{MathWorks}()]{MatlabVersion}
{MathWorks}.
\newblock {MATLAB}.
\newblock \url{https://www.mathworks.com}.
\newblock {Version: R2023a}.

\bibitem[Nasyir et~al.(2014)Nasyir, Pramujati, Nurhadi, and
  Pitowarno]{Nasyir2014}
T.~M. Nasyir, B.~Pramujati, H.~Nurhadi, and E.~Pitowarno.
\newblock Control simulation of an automatic turret gun based on force control
  method.
\newblock In \emph{2014 International Conference on Intelligent Autonomous
  Agents, Networks and Systems}. {IEEE}, aug 2014.
\newblock \doi{10.1109/inagentsys.2014.7005718}.

\bibitem[Padilla et~al.(2020)Padilla, Netto, and da~Silva]{Padilla2020Survey}
R.~Padilla, S.~L. Netto, and E.~A.~B. da~Silva.
\newblock A survey on performance metrics for object-detection algorithms.
\newblock In \emph{2020 International Conference on Systems, Signals and Image
  Processing ({IWSSIP})}. {IEEE}, July 2020.
\newblock \doi{10.1109/iwssip48289.2020.9145130}.

\bibitem[Padilla et~al.(2021)Padilla, Passos, Dias, Netto, and
  da~Silva]{Padilla2021Comparative}
R.~Padilla, W.~L. Passos, T.~L.~B. Dias, S.~L. Netto, and E.~A.~B. da~Silva.
\newblock A comparative analysis of object detection metrics with a companion
  open-source toolkit.
\newblock \emph{Electronics}, 10\penalty0 (3):\penalty0 279, Jan. 2021.
\newblock \doi{10.3390/electronics10030279}.

\bibitem[Pathak et~al.(2018)Pathak, Pandey, and
  Rautaray]{Pathak2018Application}
A.~R. Pathak, M.~Pandey, and S.~Rautaray.
\newblock Application of deep learning for object detection.
\newblock \emph{Procedia Computer Science}, 132:\penalty0 1706--1717, 2018.
\newblock \doi{10.1016/j.procs.2018.05.144}.

\bibitem[Pierucci and Bocchi(2007)]{Pierucci2007Improvements}
L.~Pierucci and L.~Bocchi.
\newblock Improvements of radar clutter classification in air traffic control
  environment.
\newblock In \emph{2007 IEEE International Symposium on Signal Processing and
  Information Technology}, pages 721--724. IEEE, Dec. 2007.
\newblock \doi{10.1109/isspit.2007.4458097}.

\bibitem[Profeta et~al.(2016)Profeta, Rodriguez, and
  Clouse]{Profeta2016Convolutional}
A.~Profeta, A.~Rodriguez, and H.~S. Clouse.
\newblock Convolutional neural networks for synthetic aperture radar
  classification.
\newblock In E.~Zelnio and F.~D. Garber, editors, \emph{SPIE Proceedings},
  volume 9843. SPIE, May 2016.
\newblock \doi{10.1117/12.2225934}.

\bibitem[Rahmat et~al.(2016)Rahmat, Hudha, Idris, and
  Amer]{rahmat-et-al-AIMT2016}
M.~S. Rahmat, K.~Hudha, A.~Idris, and N.~H. Amer.
\newblock Sliding mode control of target tracking system for two degrees of
  freedom gun turret model.
\newblock \emph{Advances in Military Technology}, 11\penalty0 (1):\penalty0
  13--28, 2016.

\bibitem[Ratches(2011)]{Ratches2011}
J.~A. Ratches.
\newblock Review of current aided/automatic target acquisition technology for
  military target acquisition tasks.
\newblock \emph{Optical Engineering}, 50\penalty0 (7):\penalty0 072001, July
  2011.
\newblock \doi{10.1117/1.3601879}.

\bibitem[Redmon et~al.(2016)Redmon, Divvala, Girshick, and
  Farhadi]{Redmon2016You}
J.~Redmon, S.~Divvala, R.~Girshick, and A.~Farhadi.
\newblock You only look once: Unified, real-time object detection.
\newblock In \emph{2016 {IEEE} Conference on Computer Vision and Pattern
  Recognition ({CVPR})}. {IEEE}, June 2016.
\newblock \doi{10.1109/cvpr.2016.91}.

\bibitem[Reiner et~al.(2017)Reiner, Hollands, and Jamieson]{Reiner2017Target}
A.~J. Reiner, J.~G. Hollands, and G.~A. Jamieson.
\newblock Target detection and identification performance using an automatic
  target detection system.
\newblock \emph{Human Factors: The Journal of the Human Factors and Ergonomics
  Society}, 59\penalty0 (2):\penalty0 242--258, Mar. 2017.
\newblock ISSN 1547-8181.
\newblock \doi{10.1177/0018720816670768}.

\bibitem[Ren et~al.(2017)Ren, He, Girshick, and Sun]{Ren2017Faster}
S.~Ren, K.~He, R.~Girshick, and J.~Sun.
\newblock Faster {R-CNN}: Towards real-time object detection with region
  proposal networks.
\newblock \emph{{IEEE} Transactions on Pattern Analysis and Machine
  Intelligence}, 39\penalty0 (6):\penalty0 1137--1149, June 2017.
\newblock \doi{10.1109/tpami.2016.2577031}.

\bibitem[Riegler et~al.(1998)Riegler, Stewart, Fitzhugh, Janson, and
  Kuperman]{Riegler1998Human}
J.~T. Riegler, R.~L. Stewart, E.~W. Fitzhugh, W.~P. Janson, and G.~G. Kuperman.
\newblock A human factors evaluation of esar/atr integration for the theater
  missile defense automatic target recognition rapid response targeting against
  mobile ground targets program.
\newblock Technical Report AFRL-HE-WP-TR-1998-0093, US Air Force Research
  Laboratory, Wright-Patterson AFB, OH, July 1998.

\bibitem[Sermanet et~al.(2014)Sermanet, Eigen, Zhang, Mathieu, Fergus, and
  LeCun]{Sermanet2014OverFeat}
P.~Sermanet, D.~Eigen, X.~Zhang, M.~Mathieu, R.~Fergus, and Y.~LeCun.
\newblock Overfeat: Integrated recognition, localization and detection using
  convolutional networks.
\newblock In \emph{International Conference on Learning Representations}.
  arXiv, Feb. 2014.
\newblock \doi{10.48550/ARXIV.1312.6229}.

\bibitem[Shetty et~al.(2021)Shetty, Saha, Sanghvi, Save, and
  Patel]{Shetty2021object}
A.~K. Shetty, I.~Saha, R.~M. Sanghvi, S.~A. Save, and Y.~J. Patel.
\newblock A review: Object detection models.
\newblock In \emph{2021 6th International Conference for Convergence in
  Technology (I2CT)}. {IEEE}, Apr. 2021.
\newblock \doi{10.1109/i2ct51068.2021.9417895}.

\bibitem[Sikka et~al.(1989)Sikka, Varshney, and
  Vannicola]{Sikka1989Distributed}
D.~I. Sikka, P.~K. Varshney, and V.~C. Vannicola.
\newblock A distributed artificial intelligence approach to object
  identification and classification.
\newblock In C.~B. Weaver, editor, \emph{Sensor Fusion II}, volume 1100, pages
  73--84. SPIE, Sept. 1989.
\newblock \doi{10.1117/12.960483}.

\bibitem[Strom(2013)]{strohm2013introduction}
L.~S. Strom.
\newblock An introduction to the sources of delivery error for direct-fire
  ballistic projectiles.
\newblock Technical Report ARL-TR-6494, Army Research Lab, Aberdeen Proving
  Ground, MD, July 2013.

\bibitem[Verly et~al.(1989)Verly, Delanoy, and Dudgeon]{Verly1989Machine}
J.~G. Verly, R.~L. Delanoy, and D.~E. Dudgeon.
\newblock Machine intelligence technology for automatic target recognition.
\newblock \emph{The Lincoln Laboratory Journal}, 2, 1989.

\bibitem[Wang et~al.(2015)Wang, Chen, Xu, and Jin]{Wang2015Application}
H.~Wang, S.~Chen, F.~Xu, and Y.~Jin.
\newblock Application of deep-learning algorithms to mstar data.
\newblock In \emph{2015 IEEE International Geoscience and Remote Sensing
  Symposium (IGARSS)}, pages 3743--3745. IEEE, July 2015.
\newblock \doi{10.1109/igarss.2015.7326637}.

\bibitem[Weaver~Jr(1990)]{weaver1990system}
J.~M. Weaver~Jr.
\newblock System error budgets, target distributions and hitting performance
  estimates for general-purpose rifles and sniper rifles of 7.62 x 51 mm and
  larger calibers.
\newblock Technical Report AD-A228-398, US Army Material Systems Analysis
  Activity, Aberdeen Proving Ground, MD, May 1990.

\bibitem[Xia et~al.(2016)Xia, Pu, Fu, and Ye]{Xia2016modeling}
Y.~Xia, F.~Pu, M.~Fu, and L.~Ye.
\newblock Modeling and compound control for unmanned turret system with
  coupling.
\newblock \emph{{IEEE} Transactions on Industrial Electronics}, 63\penalty0
  (9):\penalty0 5794--5803, Sept. 2016.
\newblock \doi{10.1109/tie.2016.2578838}.

\bibitem[Yuan et~al.(2021)Yuan, Deng, Ge, Yao, and Yang]{Yuan2021precision}
S.~Yuan, W.~Deng, Y.~Ge, J.~Yao, and G.~Yang.
\newblock Nonlinear adaptive robust precision pointing control of tank servo
  systems.
\newblock \emph{{IEEE} Access}, 9:\penalty0 23385--23397, 2021.
\newblock \doi{10.1109/access.2021.3054178}.

\bibitem[Yuan et~al.(2024)Yuan, Deng, Yao, and Yang]{Yuan2024bidirectional}
S.~Yuan, W.~Deng, J.~Yao, and G.~Yang.
\newblock Nonlinear robust adaptive control for bidirectional stabilization
  system of all-electric tank with unknown actuator backlash compensation and
  disturbance estimation.
\newblock \emph{Defence Technology}, 32:\penalty0 144--158, Feb. 2024.
\newblock ISSN 2214-9147.
\newblock \doi{10.1016/j.dt.2023.05.019}.

\end{thebibliography}

\appendix
\section{Appendix}\label{appendix}

\subsection{Equations of Motion}
\label{eqnsofmotion}
 \begin{figure}[!htb]
    \centering
    \includegraphics[scale=0.5]{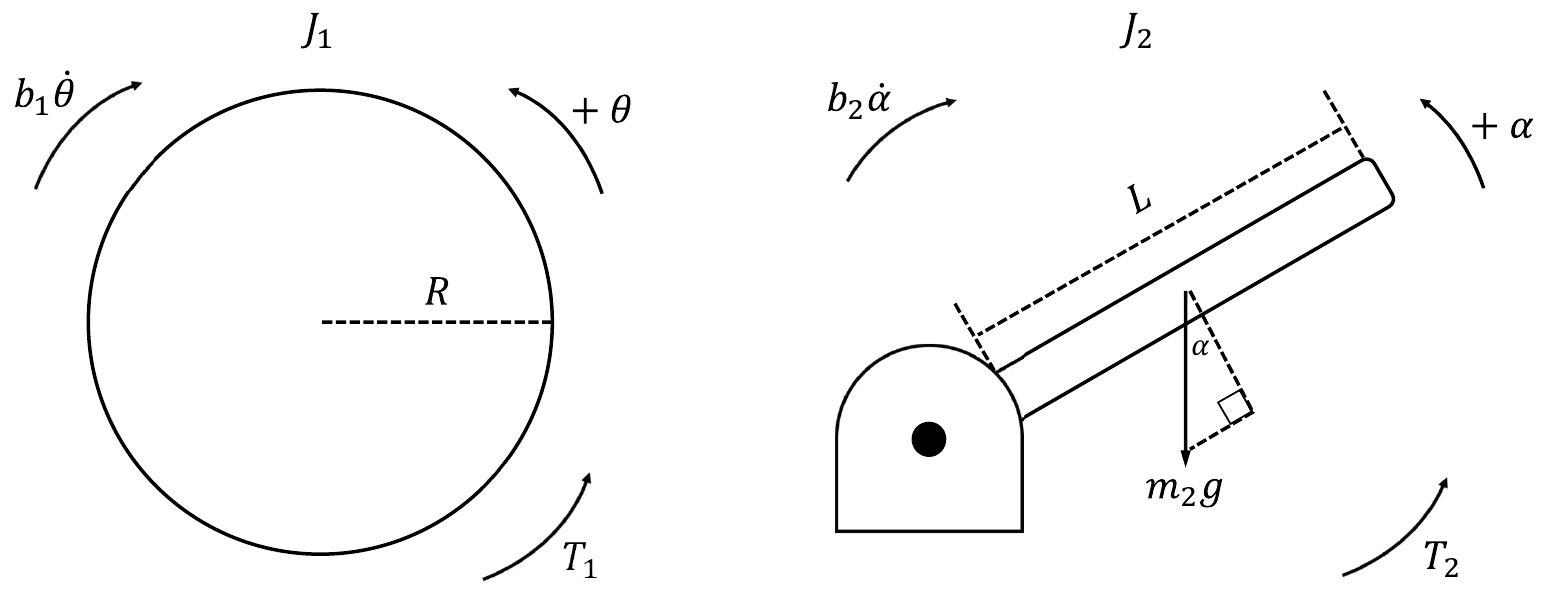}
    \caption{Free body diagram of the platform and gun barrel.}
    \label{free_body_diagram}
\end{figure}

We derive the equations of motion for the gun turret using Newton's second law for angular motion, which is
\begin{equation}
    \sum_i M_{i,O} = J\dot{\omega},
    \label{nslrotation}
\end{equation}
where $M_{i,O}$ is a moment about the point $O$ in a body with moment of inertia $J$ and $\omega$ is the angular velocity. We appeal to the free body diagram in Figure \ref{free_body_diagram} that illustrates the forces and moments acting on the system and the sign convention. Although not shown in the diagram, there is a torque acting on the gun barrel in the clockwise direction due to the weight, which has a moment arm of $L/2$. By summing the moments and applying \eqref{nslrotation}, we obtain the following two equations:
    \begin{align}
        J_1\ddot{\theta}& = -b_1\dot{\theta} + T_1 
        \label{thetaeqnofmotionderiv}\\
        J_2\ddot{\alpha}& = -b_2\dot{\alpha} - \frac{1}{2}m_2gL\cos\alpha + T_2,
        \label{alphaeqnofmotionderiv}
    \end{align}   
Moving all terms with derivatives to the left-hand side of \eqref{thetaeqnofmotionderiv} and \eqref{alphaeqnofmotionderiv} results in equations \eqref{thetaeqnofmotion} and \eqref{alphaeqnofmotion} for the gun turret outputs $\theta$ and $\alpha$. 
 
\subsection{PI+lead Controller Design}
\label{pileaddesign}

Given a desired $\omega_{gc}$ and PM:
\begin{enumerate}
    \item Calculate the magnitude and phase of the plant at $\omega_c$: $|G(j\omega_{gc})|$, $\angle G(j\omega_{gc})$.
    \item Calculate the amount of phase to be added:
    \begin{equation*}
        \phi_{\text{add}} = \text{PM} - 180^{\circ} - \angle G(j\omega_{gc}) + 6^{\circ}.
    \end{equation*}
    The extra $6^{\circ}$ accounts for the reduction in phase from PI control.
    \item Calculate $\gamma$:
    \begin{equation*}
        \gamma = \frac{1-\sin\phi_{\text{add}}}{1+\sin\phi_{\text{add}}}.
    \end{equation*}
    \item The phase is added at $\omega_{gc} = 1/\sqrt{\gamma}T_D$. Calculate $T_D$ to ensure the phase is added at $\omega_{gc}$:
    \begin{equation*}
        T_D = \frac{1}{\sqrt{\gamma}\omega_{gc}}.
    \end{equation*}
    \item Calculate $K_p$ to make $|C(j\omega_{gc})G(j\omega_{gc})|$ unity:
    \begin{equation*}
        K_P = \frac{\sqrt{\gamma}}{|G(j\omega_{gc})|}.
    \end{equation*}
    \item Make the zero $1/T_I$ a decade below $\omega_{gc}$:
    \begin{equation*}
        T_I = \frac{10}{\omega_{gc}}.
    \end{equation*}
    \item Check the response and return to step 1 if further adjustment is necessary.
\end{enumerate}

\subsection{Error Response Model and Standard Deviation Due to White Noise}
\label{errorstdevderivation}
In an ATR system, the object detection algorithm processes images of targets in real-time and the output is used to measure the aim point at each update. These images are acquired in various environmental conditions (e.g., day or night) and in different modes (e.g., RGB or thermal) that can cause image quality to fluctuate. Consequently, the alterations in image quality add uncertainty to the detections made by the algorithm that propagates through the controlled gun turret system. The resulting uncertainty added to the error in moving the gun turret to the aimpoint can be estimated by the standard deviation of the azimuth and elevation error responses at steady-state. For this purpose, we incorporate a model for the aimpoint update made by the ATR system in processing images in real-time.

At time $t\ge 0$, we assume the reference input $r(t)$ to the controlled gun turret is the response to $w(t)$, which is a random variable sampled from a normal distribution with \num{0} mean and standard deviation $\sigma_w$. The transfer function model between $w(t)$ and $r(t)$ in the Laplace domain is chosen as the following: 
\begin{equation}
    R(s) = \frac{1}{\tau s + 1}W(s),
    \label{whitenoise2ref}
\end{equation}
where $R(s)$ is the Laplace transform of $r(t)$ and $W(s)$ is the Laplace transform of $w(t)$; $s$ is the Laplace variable in units of \si{\radian/\second}. The transfer function in \eqref{whitenoise2ref} intends to model how the object detector updates the aimpoint in real-time. The time constant $\tau$ is an estimate of the speed at which the object detector processes an image and outputs bounding box locations of detected targets. We assume that this procedure occurs at \SI{0.5}{\hertz}, and accordingly, we set the time constant to $\tau = 2$ seconds. 

The transfer function in \eqref{whitenoise2ref} is used to determine the relationship between $w(t)$ and the error $e(t)$ in moving the controlled gun turret to the aimpoint. Using a block diagram of the closed-loop system, the turret movement error in the Laplace domain is
\begin{equation}
    E(s) = \frac{1}{1+K(s)G(s)}R(s),
    \label{errorsystemtf}
\end{equation}
where $K(s)$ and $G(s)$ are the controller and gun turret system transfer functions, respectively. By substituting \eqref{whitenoise2ref} into \eqref{errorsystemtf}, the transfer function between
$w(t)$ and $e(t)$ is the rational function multiplying $W(s)$ in the following:
\begin{equation}
    E(s) = \frac{1}{(2s+1)(1+K(s)G(s))}W(s).
    \label{w2errorsystem}
\end{equation}
Note there are two transfer functions in the form of the rational function in \eqref{w2errorsystem}, one each for the azimuth and elevation subsystems. 

Since the error systems are linear and the random variable $w(t)$ is normally distributed, the error responses to $w(t)$ are also normally distributed. Using results from linear system theory, the standard deviation of the azimuth and elevation error responses can be determined from the 2-norm of the error system transfer function in \eqref{w2errorsystem}. In discrete time, the error system 2-norm is defined as
\begin{equation}
    \Vert H\Vert_2 = \sqrt{\frac{1}{2\pi}\int_{-\omega_N}^{\omega_N}|H(e^{j\omega h)}|^2\,d\omega},
    \label{2norm}
\end{equation}
where $H(z)$, with $z = e^{sh}$ for a complex number $s$, is the discrete-time transfer function between the input $w(t)$ and output $e(t)$ derived from the continuous-time transfer function in \eqref{w2errorsystem}. The sample time is $h$ in units of seconds, the sample frequency is $f_s=1/h$ in units of Hertz and $\omega_N$ is the Nyquist frequency, or half the sampling frequency in units of \si{\radian/\second}. Given the white noise input $w(t)$, the standard deviation of the gun turret error response at steady-state is calculated as follows.

In the frequency domain, the variance of the white noise signal $w(t)$ \citep{Bendat2010RandomData} is
\begin{equation*}
    \sigma_w^2 = \frac{1}{2\pi}\int_{-\omega_N}^{\omega_N}S_w(\omega)\,d\omega,
    \label{whitenoisevar}
\end{equation*}
where $S_w(\omega)$ is the spectral density of $w(t)$. Since $w(t)$ is white noise, by definition $S_w(\omega)$ is constant and, therefore, can be taken out of the integral. Using the identity $\omega_N = \pi f_s$, it then follows that
\begin{align*}
    \sigma_w^2 &= \frac{1}{2\pi}S_w(\omega)\int_{-\omega_N}^{\omega_N}\,d\omega \\
    &= \frac{\omega_N}{\pi}S_w(\omega) \\
    &= f_s S_w(\omega).
\end{align*}
We then obtain the identity
\begin{equation}
    S_w(\omega) = \frac{\sigma_w^2}{f_s}.
    \label{whitenoisesd}
\end{equation}

The variance of the error response $e(t)$ to the white noise signal $w(t)$ in the frequency domain is
\begin{equation}
    \sigma_e^2 = \frac{1}{2\pi}\int_{-\omega_N}^{\omega_N}S_e(\omega)\,d\omega,
\end{equation}
where the spectral density of the error response is $S_e(\omega) = |H(e^{j\omega h})|^2S_w(\omega)$.

Using \eqref{whitenoisesd}, the variance of the error becomes
\begin{align*}
    \sigma_e^2 &= \frac{1}{2\pi}\int_{-\omega_N}^{\omega_N} |H(e^{j\omega h})|^2S_w(\omega)\,d\omega \\
    &= \frac{1}{2\pi}\int_{-\omega_N}^{\omega_N}|H(e^{j\omega h})|^2\cdot\frac{\sigma^2}{f_s}\,d\omega \\
    &= \frac{1}{2\pi}\int_{-\omega_N}^{\omega_N}|H(e^{j\omega h})|^2\,d\omega\cdot\frac{\sigma^2}{f_s}\\
    &= \Vert H\Vert_2^2\frac{\sigma_w^2}{f_s},
\end{align*}
where we have applied \eqref{2norm} to the integral term. Taking the square-root on both sides gives the standard deviation of the error response:
\begin{equation}
\label{errorStdevWhiteNoise}
\sigma_e = \Vert H\Vert_2\frac{\sigma_w}{\sqrt{f_s}}.
\end{equation}

\subsection{Variance of the Total Error}

Here we prove equation \eqref{varianceRrg} for the variance of the total error sample $\sigma_{rg}^2$. We introduce the notation $\text{E}(x)=\frac{1}{
N}\sum_{i=1}^Nx_i$ for the mean of a sample of observations $\{x_i\}_{i=1}^N$ of a random variable $x$. In this notation, the variance of the sample  is $\sigma_x^2 = \text{E}[(x-\mu_x)^2]$, where $\mu_x =\text{E}(x)$. Similarly, the covariance between two random variables $x$ and $y$ is $\text{Cov}(x,y)=\text{E}[(x-\mu_x)(y-\mu_y)]$.

Next, we will use the following properties of the sample mean, which hold for any two random variables $x$ and $y$, and any real number $a$:
\begin{equation}
    \label{meanProperties}
    \text{E}\left(x+y\right) = \text{E}(x)+\text{E}(y), \quad \text{E}(ax) = a\text{E}(x).
\end{equation}
We will also use the following identity for any real numbers $u$, $v$ and $w$:
\begin{equation}
    \label{algebraIdentity}
    (u+v+w)^2 = u^2+v^2+w^2+2uv+2uw+2vw.
\end{equation}

Now we let $\mu_{rg} = \text{E}(\Vert\mathbf{R}_{rg}\Vert)$, $\mu_{bg} = \text{E}(\Vert\mathbf{R}_{bg}\Vert)$ and $\mu_{br} = \text{E}(\Vert\mathbf{R}_{br}\Vert)$. Recall that the total error is $\Vert\mathbf{R}_{rg}\Vert = \Vert\mathbf{R}_{bg}\Vert + \Vert\mathbf{R}_{br}\Vert - \varepsilon$, where $\varepsilon$ has been defined in Section \ref{threeErrorDistributions}. It follows from \eqref{meanProperties} that the mean of the total error is $\mu_{rg}=\mu_{bg}+\mu_{br}-\mu_{\varepsilon}$. Then, by taking $u = \Vert\mathbf{R}_{bg}\Vert - \mu_{bg}$, $v = \Vert\mathbf{R}_{br}\Vert - \mu_{br}$ and $w = -(\varepsilon-\mu_{\varepsilon})$, and applying \eqref{meanProperties} and \eqref{algebraIdentity}, the variance of the total error sample becomes:
\begin{align*}
    \sigma_{rg}^2 &= \text{E}\left[\left(\Vert\mathbf{R}_{rg}\Vert-\mu_{rg}\right)^2\right] \\
    &= \text{E}\left[\left(\Vert\mathbf{R}_{bg}\Vert - \mu_{bg} + \Vert\mathbf{R}_{br}\Vert - \mu_{br} - (\varepsilon - \mu_{\varepsilon})\right)^2\right]\\
    &= \text{E}\left[\left(u + v + w
     \right)^2\right]\\
     &= \text{E}\left(u^2 + v^2 + w^2 + 2uv + 2uw + 2vw\right)
     \\ &= \text{E}(u^2) + \text{E}(v^2) + \text{E}(w^2) + 2\text{E}(uv) + 2\text{E}(uw) + 2\text{E}(vw)
      \\
     &= \sigma_{bg}^2 + \sigma_{br}^2 + \sigma_{\varepsilon}^2 + 2\text{Cov}\left(\Vert\text{R}_{bg}\Vert,\Vert\text{R}_{bg}\Vert\right) - 2\text{Cov}\left(\Vert\text{R}_{bg}\Vert,\varepsilon\right) - 2\text{Cov}\left(\Vert\text{R}_{br}\Vert,\varepsilon\right).
\end{align*}
The last identity is equation
\eqref{varianceRrg}.

\end{document}